\documentclass[aps,pra,twocolumn,floatfix,a4paper]{revtex4-1}

\usepackage{amsmath}
\usepackage{graphicx}
\usepackage{amssymb}
\usepackage{natbib}
\usepackage{hyperref}
\usepackage[T1]{fontenc}

\newcommand{\xxi}{\Xi}
\newcommand{\oldsigma}{a_z}
\newcommand{\FG}{Vertical trapping}
\newcommand{\fg}{vertical trapping}
\newcommand{\MG}{Horizontal trapping}
\newcommand{\mg}{horizontal trapping}
\newcommand{\oldGravityBoxL}{L}
\newcommand{\psii}{\Phi_n(z)}
\newcommand{\psiik}{\left|\Phi_n\right\rangle}
\newcommand{\psiicc}{\Phi_n^\ast(z)}
\newcommand{\psif}{{\Psi_{k}(z)}}
\newcommand{\psifk}{\left|\Psi_{k}\right\rangle}
\newcommand{\psifa}{\Psi_\kappa(z)}
\newcommand{\psifBC}{\Psi_\kappa(-\oldGravityBoxL)}
\newcommand{\nf}{{n}}
\newcommand{\oldalpha}{{\eta}}
\newcommand{\oldell}{\ensuremath{w}}
\newcommand{\AiryConstant}{{\cal C}}
\newcommand{\mass}{M}
\newcommand{\wrf}{\ensuremath{\omega_\text{rf}}}
\newcommand{\wosc}{\ensuremath{\omega_z}}
\newcommand{\Rzero}{\ensuremath{\Omega_0}}
\newcommand{\mfu}{m_F}
\newcommand{\mfp}{m_{\!F}'}

\begin{document}

\title{Non-adiabatic losses from radio-frequency dressed cold atom traps:\\ beyond the Landau-Zener model}

\author{Kathryn A Burrows}
\affiliation{University of Sussex, Department of Physics and Astronomy, Brighton, BN1 9QH, United Kingdom}

\author{H\'el\`ene Perrin}
\affiliation{Laboratoire de physique des lasers, CNRS, Universit\'e Paris 13, Sorbonne Paris Cit\'e, 99 avenue J.-B. Cl\'ement, F-93430 Villetaneuse, France}

\author{Barry M Garraway}
\affiliation{University of Sussex, Department of Physics and Astronomy, Brighton, BN1 9QH, United Kingdom}
\date{\today}

\begin{abstract}
  Non-adiabatic decay rates for a radio-frequency dressed magnetic
  trap are calculated using Fermi's Golden Rule: that is, we examine
  the probability for a single atom to make transitions out of the
  dressed trap and into a continuum in the adiabatic limit, where
  perturbation theory can be applied.  This approach can be compared
  to the semi-classical Landau-Zener theory of a resonant dressed atom
  trap, and it is found that, when carefully implemented, the
  Landau-Zener theory overestimates the rate of non-adiabatic spin
  flip transitions in the adiabatic limit.  This indicates that care
  is needed when determining requirements on trap Rabi frequency and
  magnetic field gradient in practical atom traps.
\end{abstract}


\maketitle


\section{Introduction}
\label{sec:introduction}

The control of ultra-cold atomic systems holds great promise for
applications in quantum technology such as sensors for gravity,
magnetism and motion. Magnetic trapping is one of the ways of both
trapping and controlling the atoms. However, by introducing
radio-frequency (rf) fields \cite{Garraway2016,Perrin2017} we can form
adiabatic, or dressed potentials, which offer a high degree of control
over trapping topology and hold promise for becoming a standard tool
for manipulating atoms and atom interferometry.  This particular type
of cold atom trap was suggested by Zobay and Garraway in 2001
\cite{Zobay2001} and first experimentally achieved in Paris in 2003
\cite{Colombe2004b,Colombe2004a}.  The trapping potential is created
by the atomic interaction with applied magnetic and rf fields such
that the potential depends on the atomic Zeeman state.

However, atoms may be completely lost from the trap if they undergo a
transition from a spin state associated with a trapping adiabatic
potential to an untrapped spin state.  In general, this can happen
because an atom travels `too fast' so that the normal adiabatic
following cannot take place. The speed of the atom could cause a
rapid change in the magnetic field amplitude or direction. Or the atom
could experience a rapid change in rf amplitude or polarisation. In
this paper we focus on quantifying the atom loss associated with
change in magnetic field amplitude. This is motivated by experiments
\cite{Colombe2004a,Schumm2005b} and the need for a greater
understanding of non-adiabatic losses as the technique of adiabatic
potentials is used in ways that require working to the limits of
adiabaticity. In particular, we will present full quantum
calculations, but because of the complexity of the problem we make a
one-dimensional model. Our full problem has gravity present, and so
the 1D model is presented for two different orientations which relate
to experiments (i.e.\ horizontal \cite{Morizot2006,Heathcote2008} and
vertical orientations \cite{Colombe2004a}). We then benchmark the 1D
model to the much simpler and widely known Landau-Zener model
\cite{Landau1932,Zener1932}, which presents a semi-classical approach.

In the related case of static magnetic field traps, loss rates have
been calculated previously \cite{Sukumar1997,Brink2006}.  However, in
the present paper a theory is presented for the rate of non-adiabatic
transitions between \emph{dressed} spin states for rf-dressed cold
atom traps.  In the following Sec.~\ref{sec:basics} we will recall the
principle of rf-dressed magnetic traps and then we give the
predictions of Landau-Zener theory in
Sec.~\ref{sec:landau-zener-theory}. Section~\ref{sec:quantum-dynamics-1d-trap}
presents the quantum treatment of non adiabatic losses from the trap,
which is compared to the predictions of Landau-Zener theory in
Sec.~\ref{sec:comp-with-LZ}. Finally, we conclude the paper in
Sec.~\ref{sec:conclusion}.

\section{Basics of adiabatic potentials}
\label{sec:basics}
A single non-relativistic atom with mass $\mass$ is trapped in the $z$
direction by an adiabatic potential arising from its interaction with
two fields: a static magnetic field and an rf field.  The total
Hamiltonian describing this problem is given by
\cite{Fortagh2007,Garraway2016,Perrin2017}
\begin{equation}
\hat{H}_\mathrm{tot} = \frac{{\hat{p}_z}^2}{2\mass} 
+ g_F \frac{\mu_B}{\hbar} \mathbf{F} \cdot \mathbf{B}(\hat{z}) 
+ g_F \frac{\mu_B}{\hbar} \mathbf{F} \cdot \mathbf{B}_\mathrm{rf}(t) 
+ \mass g  \hat{z} \,.
\label{eq:H-in-Intro}
\end{equation}
This forms the 1D model which we study.  The first term in the
Hamiltonian~(\ref{eq:H-in-Intro}) is the atom's kinetic energy, where
$ {\hat{p}_z} $ is the momentum operator in the $z$ direction. The
second term describes how the atom responds to a static magnetic field
$ \mathbf{B}(z)$.  The total angular momentum of the atom is
$\mathbf{F}$ in multiples of $ \hbar$, $\mu_B $ is the Bohr magneton,
and $ g_F $ is the Land\'e factor.  The third term in
Eq.~(\ref{eq:H-in-Intro}) represents the corresponding interaction
with a uniform rf field, and the last term is the gravitational
potential of the atom where $g$ is the gravitational acceleration.
This last term may be present, or not, depending on the orientation of
the direction of 1D trapping with respect to local gravity. We will
consider two particular cases: in the case where the trapping
direction is perpendicular to local gravity we will refer to a \mg\
model, where $g=0$ in Eq.~(\ref{eq:H-in-Intro}); otherwise we refer to
a \fg\ model (where the direction of motion will still be $z$). By
considering a purely one-dimensional model we will be neglecting the
possibility for a change in orbital angular momentum in the trap,
which could be possible if motion in other directions is also included
(for the pure magnetic trap case, see
Refs.~\cite{Sukumar1997,Brink2006}).

An rf-dressed adiabatic potential results from the Hamiltonian of
Eq.~(\ref{eq:H-in-Intro}) when either the static or the rf field (or
both) vary with position
\cite{Garraway2016,Perrin2017,Zobay2001,Zobay2004}.  In this paper we
consider a static magnetic field of the form $\mathbf{B}(z) = B(z)
\mathbf{e}_z$, which can be found, for example, on-axis in a
quadrupole field. The direction of this field is fixed, but the
variation of the magnitude of the field with $z$ plays a crucial role
in this paper.  We take the radio-frequency (rf) magnetic field to
have the form $\mathbf{B}_\mathrm{rf}(t) = B_\mathrm{rf} \cos({\wrf\,
  t}) \mathbf{e}_x $, which gives maximal coupling for a linearly
polarized field. Other polarisations are possible without significant
change to the details below \cite{Perrin2017}. The applied rf field
induces the atoms to undergo transitions between the different Zeeman
states within a single hyperfine spin manifold such that the atoms are
confined and forced to oscillate near the location where the frequency
of the applied rf field matches the frequency splitting of the Zeeman
sub-levels \cite{Garraway2016,Perrin2017}.

In the analysis here the rf field is treated classically and
$B_\mathrm{rf}$ is position independent (which is suitable for rf
fields generated by macroscopic coils, but not generally suitable for
atom chip cases where the rf field is generated `on-chip'
\cite{Schumm2005b,ReichelVuletic}). To obtain the adiabatic
potentials we first utilize a unitary transformation
\begin{equation}
\hat{U}_1 = \exp\left(-i s
  \frac{\wrf t}{\hbar}  \hat{F}_z\right) , 
\label{eq:U1}
\end{equation}
where the quantity $s$ represents the sign of $g_F$, i.e.
\begin{equation}
s = \frac{g_F}{\left| g_F \right|}
\,.
\label{eq:def:s}
\end{equation}
Using this spin rotation to change our basis we obtain a transformed
Hamiltonian via $ \hat H^\prime_\text{tot} = \hat{U}^\dagger_1 \hat
H_\text{tot} \hat{U}_1 - i \hbar \hat{U}_1^\dagger \partial_t
\hat{U}_1 $.  Then, after also making the rotating wave approximation
(RWA) we find \cite{Garraway2016,Perrin2017}
\begin{equation}
\hat H^\prime_\text{tot} 
  =  \frac{{\hat{p}_z}^2}{2\mass} 
   +  \hat{H}_\mathrm{RWA} 
        + \mass g \hat z ,
\label{LZHamiltonian1}
\end{equation}
where 
\begin{equation}
  \hat{H}_\mathrm{RWA} 
  = 
     s \left[ -  \delta(\hat z) \hat{F}_z +  \Rzero \hat{F}_x  \right]
\,.
\label{LZHamiltonian2}
\end{equation}
We note that, in anticipation of
Sec.~\ref{sec:quantum-dynamics-1d-trap}, the unitary transformation
$\hat{U}_1$ commutes with the momentum operator $\hat p_z$, which
leaves the kinetic operator unchanged in Eq.~\eqref{LZHamiltonian1}.
The Rabi frequency $\Rzero$ is used as a measure of the strength of
the coupling between the rf field and the atom and is given by $\Rzero
= \left| g_F \right| \frac{\mu_B}{2\hbar}
B_\mathrm{rf} \label{RabiFrequency} $.  The detuning, i.e.\ the
frequency difference between the Zeeman split energy levels at $z$ and
the rf field frequency of oscillation, is given by
\begin{equation}
\delta(z) = \wrf - \frac{\left| g_F\mu_BB(z)\right|}{\hbar} .
\label{delta}
\end{equation}

\begin{figure}[ht]
\includegraphics[width=\linewidth]{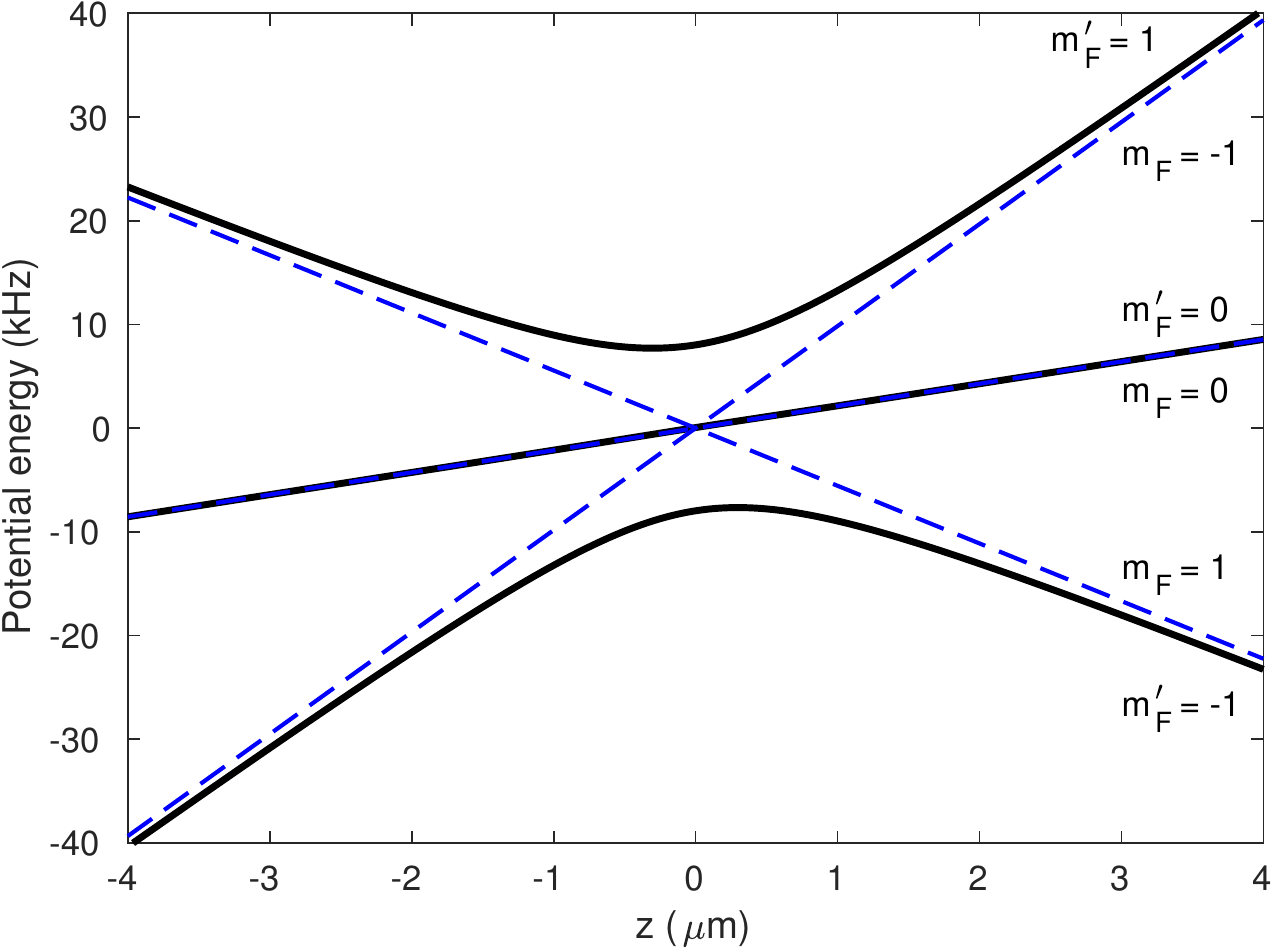}
\caption[Energy level diagram for rf-dressed trap]{Bare 
and adiabatic potentials $V$ as a function of atomic position $z$
for an atom with total angular momentum $F = 1$ in an rf-dressed cold
atom trap. The gravitational potential is included and results in the
slope of the coincident potentials labelled $\mfu=0$ and
$\mfp=0$. Blue dashed lines show the potentials given in
Eq.~(\ref{eq:LZBare}), with the crossing point necessary for
Landau-Zener theory, and labelled $\mfu=-1,0,1$.  The solid black
lines show the adiabatic potentials in the dressed state basis given
by Eq.~(\ref{eq:bmg-adia-pots}) and labelled $\mfp=-1,0,1$.  To give a
concrete example, we use values taken from Ref.\cite{Merloti2013a},
i.e.\ for a ${}^{87}$Rb atom and a magnetic field gradient of
$B^\prime = 1.1$~T/m resulting in a detuning which varies linearly
with position. The Rabi frequency is set to $\Rzero / 2\pi = 8$~kHz
for this graph.
\label{fig:EZRF}}
\end{figure}

We will pay attention to the kinetic term in
Eq.~(\ref{LZHamiltonian1}) in
Sec.~\ref{sec:quant-mech-non-adia-decay-rates}. If we, for now,
neglect the kinetic term, and consider the absence of any coupling,
i.e.\ $\Rzero\longrightarrow 0$, we obtain from
Eq.~(\ref{LZHamiltonian1}) the uncoupled, or `bare' potentials
\begin{equation}
  \label{eq:LZBare}
  - \mfu s \hbar \delta(z)   + \mass g z .
\end{equation}
In the presence of a coupling $\Rzero$, but still neglecting the
kinetic term, the Hamiltonian~(\ref{LZHamiltonian1}) can be
diagonalised at any given position $z$.  We use a time-independent
spin rotation $\hat{U}_2$ about the $y$ axis with an angle $\theta$,
so that \cite{Perrin2017}
\begin{equation}
\hat{U}_2 = \exp\left({-i\theta(\hat{z})\hat{F}_y / \hbar }\right),
\label{eq:U2-defn}
\end{equation}
where
\begin{equation}
     \theta(z) = \arccos\left(-\frac{\delta(z)}{\sqrt{\delta(z)^2 +
         \Rzero^2}}\right) 
+  \frac{s-1}{2}\pi
\label{eq:def-theta}
\end{equation}
and $s$ is the sign introduced in Eq.~\eqref{eq:def:s}.  Thus we
obtain for the Hamiltonian in the adiabatic approximation
\begin{eqnarray}
  \hat H_{\rm eff} &=&     
  \hat{U}^\dagger_2  \hat H_\text{RWA} \hat{U}_2  + M g \hat{z} \nonumber\\
  &=&
  \sqrt{ \delta(\hat{z})^2 + \Rzero^2     }  \,     \hat F_z  + \mass g \hat{z} 
  \label{eq:bmg-diagonal}
\end{eqnarray}
which leads to the adiabatic potentials
\begin{equation}
  \label{eq:bmg-adia-pots}
  V_{\mfp}(z) = \mfp \hbar \sqrt{ \delta^2(z) + \Rzero^2     }   + \mass g z \,,
\end{equation}
which are trapping potentials for $\mfp>0$.  These potentials form our
underlying atom trap, and are illustrated in Fig.~\ref{fig:EZRF} for
an example with $\delta$ varying linearly in space.  To understand the
process of decay from these traps we must account for the kinetic term
neglected to reach Eq.~(\ref{eq:bmg-adia-pots}).  We will do this in
Sec.~\ref{sec:quantum-dynamics-1d-trap}.  However, we will first look
at the semi-classical Landau-Zener analysis of the situation.

\section{Landau-Zener theory}
\label{sec:landau-zener-theory}

The original Landau-Zener model \cite{Landau1932,Zener1932} is a
\emph{two}-level time-dependent model in which the coupling between
the two levels is constant and the time-dependent bare potentials
change linearly in time. We can apply it to the simplified situation
of Fig.~\ref{fig:EZRF} by assuming that the spatial potential
$\hbar\delta(z)$ is linearised about the resonance location at $z=0$
so that
\begin{equation}
\delta(z)\sim \delta'(0) z = \pm  \alpha z \,,
\label{eq:delta-linearised}
\end{equation}
where the prime indicates differentiation with respect to $z$ and the
`$\pm$' is to account for a different sign of the gradient of
$\delta(z)$ whilst keeping the magnitude of the gradient, $\alpha$,
positive so that $\alpha = |\delta'(0)|$.  A constant gradient is, for
example, exactly the situation on-axis in a quadrupole trap
\cite{Merloti2013a,Fortagh2007} where $ \alpha = | g_F\mu_B B^\prime
|/\hbar$, with $B^\prime = \frac{\partial B}{\partial z}$ being the
magnetic field gradient in the $z$ direction.

The Landau-Zener model assumes that the atom travels at constant speed
on a linear potential, at least for the duration of the crucial region
around $\delta=0$ (the `crossing point' of the `bare' states in
Fig.~\ref{fig:EZRF} where there is magnetic resonance).  These are
very crude approximations, but the model then yields the probability
for an atom to make a transition out of the adiabatic state it would
be following when it is away from the resonance location.  The
Landau-Zener model is commonly used to estimate the significance of
non-adiabatic losses from rf-dressed cold atom traps
\cite{Zobay2004,Morizot2007} by combining the transition probability
with the number of crossings per unit time due to the atom oscillating
in the adiabatic trap.  For the Landau-Zener model, we again neglect
the kinetic term in Eq.~(\ref{LZHamiltonian1}) and taking $F =
\frac{1}{2}$, together with a speed $v$ such that $z(t)\sim v t$, we
obtain
\begin{equation}
	H_\text{LZ} = 
	\frac{\hbar}{2}
\begin{pmatrix}
	 \alpha v t  & \Rzero \\
	 \Rzero       & - \alpha v t 
\end{pmatrix}
+ \mass g  v t .
\label{eq:HLZ}
\end{equation}
Following the Landau-Zener model
\cite{Landau1932,Zener1932,Child1974,Suominen1991,Garraway1991,Garraway1995,Vitanov1996,Garraway2016},
the probability for remaining in the adiabatic state is given by
\begin{equation}
	P_{LZ}\left(v\right) = 1 - \exp \left( - \frac{\pi \Rzero^2}{2
            \alpha  v} \right). \label{LZspinhalf}
\end{equation}
Gravity does not play a role in this Landau-Zener model as it is
assumed that the atom passes through the crossing point at $z=0$ and
the factor $\mass g v t $ just introduces a global phase factor. In
the original two-level model there is also an assumption that the atom
does not return through the crossing. If it does, there can be
interference effects due to the differing phase factors at the
crossing \cite{Garraway1992,Garraway1997}. As we will be working in
the adiabatic limit, we will neglect these phase factors, even though
the atom will be, in reality, oscillating in the adiabatic potential.

The standard Landau-Zener model only considers crossings between two
energy levels, however, Vitanov and Suominen \cite{Vitanov1997} have
extended the model to account for a crossing involving $2F + 1$ energy
levels.  If we consider first a single pass of the atom through the
crossing region, the probability that an atom remains in the initial
extremal adiabatic state is given by 
$ P_{LZ}^F(v) = {\left[ P_{LZ}\left(v\right) \right]}^{2F} $
\cite{Vitanov1997}. We now repeat the argument: every time the atom
traverses the crossing, the probability of being lost from the initial
adiabatic state is $1 - P_{LZ}^F$.  Thus, to obtain an estimate of the
decay rate it is necessary only to consider how many times the atom
will `pass' the crossing region per unit of time, taking into account
that the atom transverses the crossing region twice per period. The
decay rate from the Landau-Zener model as a function of atomic speed
$v$ is then given by
\begin{equation}
	\Gamma^{LZ}(v) = \frac{\wosc}{\pi} \left[ 1 - {\left[ 1 - \exp
              \left( {- \frac{\pi \Rzero^2}{2 \alpha  v}} \right) \right]}^{2F} \right], \label{LZgeneral}
\end{equation} 
where $\wosc$ is the oscillation frequency in the dressed trap. This
can be estimated from the classical motion of a particle in the
potential. For a trap which already has a sufficiently strong
coupling to be approximately adiabatic, the exponential term is very
small and we then obtain the very small decay rate
\begin{equation}
  \label{eq:LZ-adiabatic-limit}
  	\Gamma^{LZ}(v) \approx
        \frac{2 \wosc F}{\pi} \exp\left(
          -\frac{\pi\Rzero^2}{2 \alpha  v} \right).
\end{equation}
Noting that the speed $v$ in Eq.~(\ref{eq:HLZ}) is defined on the bare
potentials, we can write Eq.~(\ref{eq:LZ-adiabatic-limit}) in terms of
the approximate total energy $E=\frac{1}{2}Mv^2$ of the atom in the
bare state referenced to zero potential at the crossing, i.e.\ as
\begin{equation}
  \label{eq:LZ-adiabatic-limit2}
  	\Gamma^{LZ}(E) \approx
        \frac{2 \wosc F}{\pi} \exp\left(
          -\frac{\pi\Rzero^2}{2 \alpha } 
         \sqrt{\frac{\mass}{2E}}
        \right).
\end{equation}
This suggests that for adiabatic trapping a strong Rabi frequency
$\Rzero$ is desirable, as is a low gradient $\alpha$ and low energies
$E$.

\section{Quantum dynamics in the 1D trap}
\label{sec:quantum-dynamics-1d-trap}

To perform a quantum mechanical analysis of the
Hamiltonian~(\ref{LZHamiltonian1}) and decay from the adiabatic trap
we again approximately diagonalise Eq.~(\ref{LZHamiltonian1}), this
time including the kinetic term. We again use the rotation
$\hat{U}_2$, Eq.~(\ref{eq:U2-defn}), and we note that the position
dependence of the angle $\theta(z)$, Eq.~(\ref{eq:def-theta}),
prevents the unitary transformation $\hat{U}_2$ from commuting with
the momentum operator. The origin of this is the spatial dependence of
the static field amplitude $B(z)$. Thus, to determine the effect of
the unitary transformation $\hat{U}_2$ on the
Hamiltonian~(\ref{LZHamiltonian1}) we will need to use the relation $
\hat{U}_2^\dagger \hat{p}_z {\hat{U}_2} = \hat{p}_z - \theta'(\hat{z})
\hat{F}_y$. As a result, we find that the Hamiltonian for a single
atom, already loaded into an rf-dressed cold atom trap, can be
expressed by
\begin{equation}
	\hat{H} = \frac{{\hat{p}_z}^2}{2\mass} + \hat{V}_A \hat{F}_y +
        \hat{V}_B {\hat{F}_y}^2 + 
         \sqrt{\Rzero^2 + {\delta(\hat{z})}^2} \hat{F}_z + \mass g \hat{z}  \label{Hamiltonian}
\end{equation}
where 
\begin{eqnarray}
	\hat{V}_A &=&  
                 -\frac{1}{2\mass}\left( 2 \hat{\theta}' \hat p_z -i \hbar \hat{\theta}''
                 \right)  
\rightarrow 
\frac{i\hbar}{2\mass}\left( 2 \theta'
          \frac{\partial}{\partial z}  +  \theta''     \right)   
\nonumber\\
           &=& \frac{i \hbar}{2\mass}\left[ \frac{2 \delta {\delta^\prime}^2
               \Rzero}{{\left( \Rzero^2 + \delta^2 \right)}^2} -
             \frac{\delta^{\prime \prime} \Rzero}{\Rzero^2 + \delta^2} -
             \frac{2 \delta^\prime \Rzero}{\Rzero^2 + \delta^2}  
           \frac{\partial}{\partial z}\right]	\label{VAdelta}
\end{eqnarray}
and
\begin{equation}
	\hat{V}_B 
  = \frac{(\hat{\theta}')^2}{2\mass} 
~\rightarrow~ 
    \frac{1}{2\mass} \frac{{\delta^\prime}^2 \Rzero^2}{{\left( \Rzero^2 + \delta^2 \right)}^2}	\label{VBdelta}
\end{equation}
with a prime indicating differentiation by $z$
and with $\hat{\theta}' \equiv \theta'(\hat{z})$.

The gauge potential terms given by $\hat{V}_A \hat{F}_y$ and
$\hat{V}_B {\hat{F}_y}^2$ are often neglected to consider the
Hamiltonian in the adiabatic approximation. Here we shall use the
non-adiabatic Hamiltonian to model the losses from an rf-dressed trap
caused by transitions between dressed spin states.
Equation~(\ref{Hamiltonian}) applies for general $\delta(z)$, but in
the case where $\delta(z)$ is linearised, as in
Eq.~(\ref{eq:delta-linearised}), the first order derivative is
constant and the second order $\delta''$ is zero. In this case we can
see from Eqs.~(\ref{VAdelta}) and (\ref{VBdelta}) that $\hat{V}_A$ is
an odd function of $z$ and $\hat{V}_B$ is an even function of $z$. As
a result, in the treatment below, $\hat{V}_A$ will couple states of
opposite parity and $\hat{V}_B$ will couple states of the same parity.
Expressing the Hamiltonian as
\begin{eqnarray}
	&&\hat{H} = \frac{{\hat{p}_z}^2}{2\mass} 
          + 
         \sqrt{\Rzero^2 + {\delta(\hat{z})}^2} \hat{F}_z + \mass g \hat{z}\label{FHamiltonian}\\
         &&+ \frac{\hat{V}_A}{2i} (\hat F_+ - \hat F_-)       
         + \frac{\hat{V}_B}{2} \left( {\hat{F}}^2 - {\hat{F}_z}^2 \right) -
         \frac{\hat{V}_B}{4} \left( {\hat{F}_+}^2 + {\hat{F}_-}^2\right)\nonumber
\end{eqnarray}
where $\hat{F}_\pm = \hat{F}_x \pm i \hat{F}_y$ and ${\hat{F}}^2 =
{\hat{F}_x}^2 + {\hat{F}_y}^2 + {\hat{F}_z}^2$, it can be seen that
$\hat V_A$ gives the coupling between states with $| \Delta \mfp | =
1$ and $\hat V_B$ gives both an energy shift and the coupling between
states with $| \Delta \mfp | = 2$. For an $F = 1$ system, as displayed
in Fig.~\ref{fig:EZRF} with the trapping potential defined as the case
$\mfp = 1$, the $\hat V_A$ coupling then induces transitions to the
$\mfp = 0$ spin state and the $\hat V_B$ coupling induces transitions
to the $\mfp = -1$ spin state. Once in the $\mfp = 0$ or $\mfp = -1$
states, the atoms are highly likely to travel out of the trapping
region and be permanently lost from the trap.

\subsection{Quantum mechanical non-adiabatic decay rates}
\label{sec:quant-mech-non-adia-decay-rates}

In this section formulae for the rate of non-adiabatic spin `flips'
out of a rf-dressed cold atom trap are obtained using Fermi's Golden
Rule. This is justified in the situation where the non-adiabatic
effects act as a perturbation on the adiabatic states.

In the following development we consider two cases: first the \mg\
model (Sec.~\ref{sec:high-gradient-model}) where the orientation of
the motion, perpendicular to gravity, ensures that gravity plays no
role in the dynamics. Secondly, we consider in
Sec.~\ref{sec:low-gradient-model} a \fg\ model where gravity acts to
pull atoms out of the region of rf resonance and to modify their
oscillation frequency.  Interactions between the atoms are not
considered, making this analysis unsuitable for Bose-Einstein
condensates but reasonable for dilute atomic clouds comprised of
thermal atoms. In the following analytic development, for simplicity,
we also neglect the $\hat V_B$ coupling term as numerical
investigations have shown it to have a small effect for the parameters
of interest.  However, some of the numerical results presented do
include a contribution to the $\mfp=0, 1$ potentials from the $\hat
V_B$ term: this contribution, a non-adiabatic potential, is described
in the Appendix.

We consider a trapped atom with $F = 1$ and calculate decay rates for
the rate of transitions from the $\mfp = 1$ dressed spin state to the
$\mfp = 0$ dressed spin state.  With these assumptions we use
perturbation theory to derive equations which model the rate of
transitions between dressed spin states in an rf-dressed cold atom
trap.  Our analysis can be extended to other spin systems, but note
that for $F > 1$ transitions from the extremal trapping potential
($\mfp=F$) would not be to a continuum, as in the case $F=1$, but to
$\mfp=F-1$ which will have discrete states.

Thus the unperturbed Hamiltonian for the system is taken to be 
\begin{equation*}
\hat{H_0} =
\frac{{\hat{p}_z}^2}{2\mass} + 
\sqrt{\Rzero^2   + {\delta(\hat{z})}^2} \hat{F}_z + \mass g \hat{z} 
\end{equation*}
with the perturbing term given by $ \Delta \hat{H} = \hat{V}_A
\hat{F}_y$. The initial trapping potential is then given by $ V_i (z)
= \hbar \sqrt{\Rzero^2 + \delta(z)^2} + \mass g z $ (where $g=0$ in
the \mg\ model).  For the untrapped $\mfp = 0$ dressed spin state the
final adiabatic potential will be set to $V_f (z) = \mass g z$ (where
again $g=0$ in the \mg\ model).  The origin of the $z$ axis is the
resonance location, with $\delta(z{=}0) = 0$.

For the next sections we define the following notation where we use
the product state $ | F = 1 , \mfp = 1 \rangle \cdot \psiik $ for the
$n$th eigenfunction of a trapped atom which is composed of spin states
$| F = 1 , \mfp = 1 \rangle$ and spatial states $\psiik$ such that the
spatial wave-function $\psii$ is given by $\psii = \langle z \psiik$.
Similarly, for the un-trapped spin state we will use the product state
$ | F= 1 , \mfp = 0 \rangle \cdot \psifk $ where the label $k$ will be
associated with the outgoing momentum of the escaping atom and the
wave-function $\psif$ will be given by $\psif = \langle z \psifk$.

\subsection{\MG\ model}
\label{sec:high-gradient-model}

We first consider a situation where there is in effect no gravitational
potential, such as when the trapping is in the horizontal direction. This
is the case in a ring trap \cite{Morizot2006,Heathcote2008} where rf
adiabatic potentials provide horizontal confinement and optical light
shifts provide vertical confinement. The situation could also arise in a
micro-gravity environment, or when gravity is compensated with other fields.

A harmonic approximation of the potential $ \hbar \sqrt{ \delta(z)^2 +
  \Omega_0^2 } $ for the trapped $ \mfp = 1 $ state can be obtained by
Taylor expansion to give
\begin{equation}
	V_i(z) = \hbar \Rzero 
     + \frac{1}{2} \mass \wosc^2 z^2 
    \label{HOMG}
\end{equation}
with a trapping frequency 
\begin{equation}
	 \wosc = \alpha \sqrt{\frac{\hbar}{\mass \Rzero}}. 
\label{omega}
\end{equation}
Here we have used Eq.~(\ref{eq:delta-linearised}),
$\delta(z)=\pm\alpha z$, with positive $\alpha$, and we expect the
expansion to be valid in the region $ |z| \ll \oldell$, where
$\oldell$ characterises the range of the rf interaction and is given
by $\oldell = \Rzero / \alpha $.  The $n$th wavefunction for an atom
in the initial trapped $\mfp = 1$ spin state is given by the usual
harmonic oscillator wavefunction
\begin{equation}
\psii = 
\frac{H_{n}(z/\oldsigma)}
{\sqrt{n! 2^{n}\oldsigma\sqrt{\pi}}}\,e^{-z^2/(2 \oldsigma^2)},
\label{eq:harmWF}
\end{equation}
with the associated energy $ E_{n} = \left( n + \frac{1}{2} \right)
\hbar \wosc + \hbar \Rzero$.  $H_n$ is the Hermite polynomial of
degree $n$, and the variable $ \oldsigma = \sqrt{\hbar/(\mass \wosc)}
$ is the standard length scale associated with the harmonic oscillator
frequency $\wosc$.

For our untrapped $ \mfp = 0 $ state, in the absence of gravity
the potential for the atoms is zero, i.e.
\begin{equation}
   V_f(z) = 0.
\label{Vuntrapped}
\end{equation}
However, to allow a calculation of the density of states, we consider
the system to be confined to a region $ -L/2 < z < L/2 $, later taking
$L \rightarrow \infty$.  Starting from the harmonic state $\psiik$,
with a parity $(-1)^n$ set by the index $n$ of the harmonic oscillator
eigen-function, the wave-function $\psif$ of the final state is then
given by
\begin{equation}
 \psif = \frac{1}{\sqrt{2 L}} \left[ e^{i k z} - (-1)^{\nf} e^{-i k z} \right]
  \label{BoxWF}
\end{equation}
where $k$ stands for $k(n)$ and depends on the initial state.  The
factor $(-1)^n$ ensures that the final state and the initial state
coupled by the operator $\hat V_A$ have opposite parity as discussed
in Sec.~\ref{sec:quantum-dynamics-1d-trap}. When we apply Fermi's
Golden rule the energy $ E_{k(n)} = \hbar^2k^2(n)/(2 \mass) $ must
match the harmonic oscillator energy of the initial state, i.e.\ $
E_{n} = \left( n + \frac{1}{2} \right) \hbar \wosc + \hbar \Rzero$.
Thus we have
\begin{equation}
k(n) \oldsigma = \sqrt{1 + 2n + 2 \Rzero/\wosc},
\end{equation}
where $n$ corresponds to the index of the initial state $\psiik$.
In the following we introduce  $q(n)$, a scaled momentum, which stands for 
\begin{equation}
q(n)= k(n) \oldsigma =\sqrt{1 + 2n + 2 \oldalpha^2},
\label{eq:q.def}
\end{equation}
with the dimensionless variable $\oldalpha = \oldell/\oldsigma$ being
the Landau-Zener crossing length-scale $\oldell$ scaled to
the harmonic oscillator length-scale $\oldsigma$. Its expression as a function of $\Omega_0$ and $\alpha$ reads
\begin{equation}
\eta = \left(\frac{M}{\hbar}\right)^{\!1/4}\,\frac{\Rzero^{3/4}}{\alpha^{1/2}}\, .
\label{eq:eta}
\end{equation}

\begin{figure}[t]
\includegraphics[width=\linewidth]{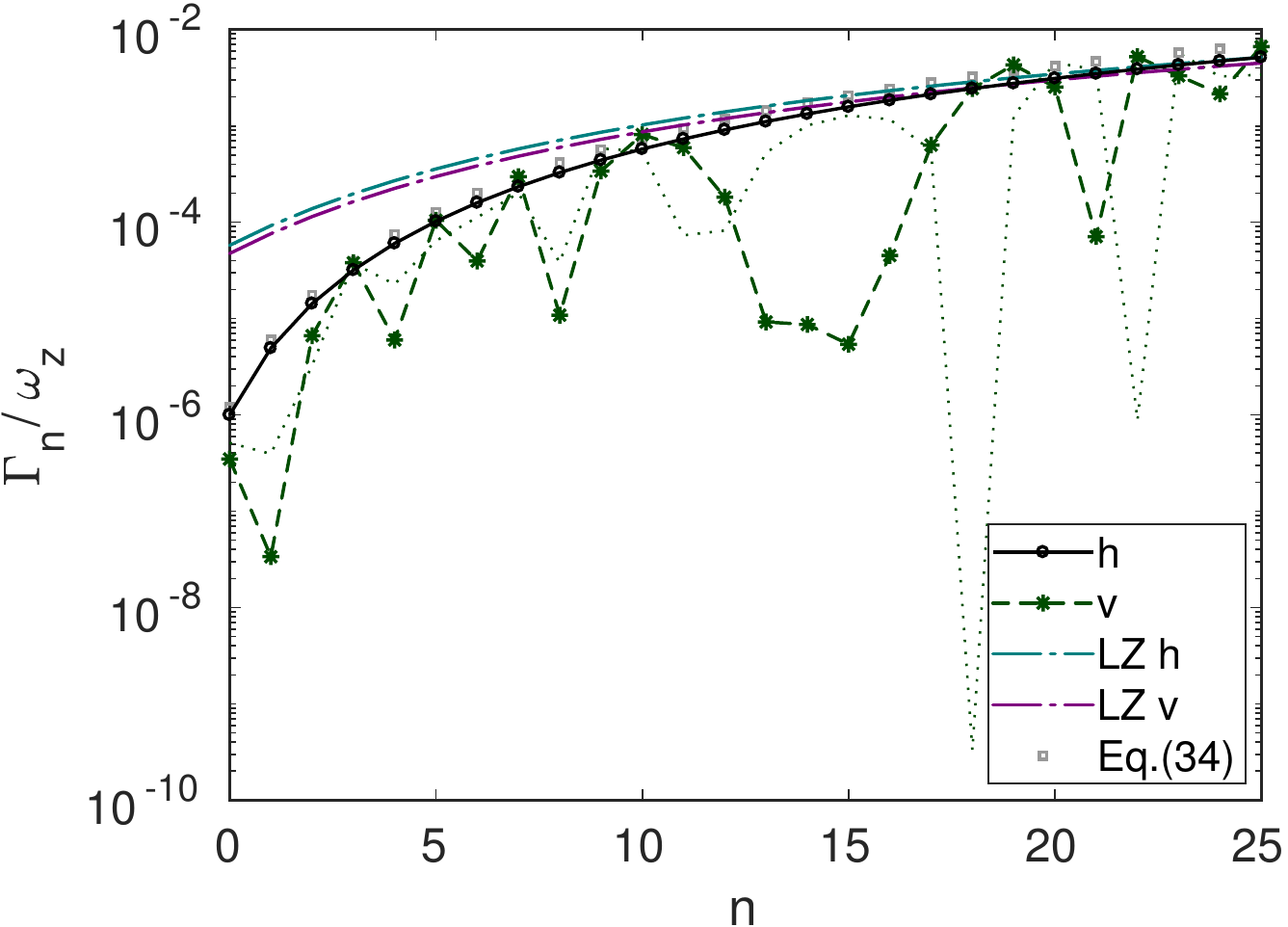}
\caption[Decay rate variation with $n$]{%
The scaled decay rate $\Gamma_n/\wosc$ is shown as a function of the
  harmonic oscillator label $n$ for different models.  The solid line
  with circles shows the result for the \mg\ model (h) as given by
  Eq.~\eqref{IntDecayMG}.  The dashed line with asterisks indicates
  the scaled \fg\ decay rates (v) as given by
  Eq.~\eqref{IntDecayFG}. These \fg\ decay rates are not a smooth
  function of $n$.  We can also include non-adiabatic potentials based
  on including part of $\hat V_B$ and resulting in modified parameters
  given in the Appendix. For the vertical trapping model, the effect
  is a shift of the location of the `dips' (dotted line). However, in
  the case of the \mg\ model, the inclusion of these corrections
  produces no visible change to the solid (h) line and is not shown.
  The chained lines (LZ~h/LZ~v) indicate the result of a Landau-Zener
  calculation given by Eq.~\eqref{LZDecay} (with $\epsilon$ set to
  zero in the `h' case).  For the parameters used (i.e.\ from
  Fig.~\ref{fig:EZRF}) the trap frequency was $\wosc/2\pi=0.93$~kHz
  for the \mg\ model (Eq.~\eqref{omega}) and corresponding
  Landau-Zener model.  For the case of the \fg\ model and its
  corresponding Landau-Zener curve, the trap frequency is found from
  Eq.~\eqref{omegag} to be $\wosc/2\pi=0.87$~kHz.  The points marked
  with small squares are given by the analytic approximation
  Eq.~\eqref{Gamman}.  The calculations are done for the $F=1$
  hyperfine ground state of $^{87}$Rb with the parameters given in
  Fig.~\ref{fig:EZRF}. This corresponds to $\oldalpha =
  \Rzero/(\alpha\oldsigma)\sim 2.9 $ for the \mg\ model, and for the
  \fg\ model: $\oldalpha \sim 2.8 $ (Eq.~(\ref{eq:def_eta})) and
  $\epsilon = \mass g / (\hbar \alpha) = 0.28$.\label{fig:fig2}}
\end{figure}

To apply Fermi's Golden rule we need the density of states 
$D(E) =  \frac{\partial N}{\partial E}$ which is given by 
\begin{equation}
D(E) =\frac{1}{2} \frac{L}{\pi\hbar}\sqrt{\frac{\mass}{2E}}
 = \frac{\mass L}{2 \pi \hbar^2 k(n)},
\end{equation}
where a factor of a half arises because we only select states with
appropriate parity.  Then, putting this all together, the Fermi's
Golden Rule decay rate for an atom with initial state $\psiik$ with
energy $E_n$ in the \mg\ model is
\begin{widetext}
\begin{eqnarray}
	\Gamma_n &=& \left|\left\langle \mfp{=}1 \right| \hat F_y \left| \mfp{=}0  \right\rangle\right|^2
	 \times \lim_{L\to\infty} \frac{\mass L}{2 \pi \hbar^2 k} \frac{2\pi}{\hbar} 
	 \left|  \int_{ -
            \frac{L}{2}}^{\frac{L}{2}} \psiicc \hat V_A \psif dz \right|^2
   \nonumber\\
&=&\wosc \frac{{\oldell}^2  \oldsigma}{2^{n+2} n! k \sqrt{\pi}} \nonumber
	\times \bigg| \bigg\{ \int_{ - \infty}^{\infty}\frac{z H_n\left(z/\oldsigma\right) e^{-\frac{z^2}{2\oldsigma^2}} }{{(z^2 + {\oldell}^2)}^2} \bigg[ e^{i k z} + (-1)^{n+ 1} e^{-i k z} \bigg] dz \nonumber
	\\ &&
	- i k \int_{ -\infty}^{\infty} \frac{ H_n\left(z/\oldsigma\right) e^{-\frac{z^2}{2\oldsigma^2}}}{z^2 + {\oldell}^2} \bigg[ e^{i k z} + (-1)^{n} e^{-i k z} \bigg] dz \bigg\} \bigg|^2, \label{IntDecayMG1}
\end{eqnarray}
where we have used 
$|\langle \mfp{=}1 | \hat F_y | \mfp{=}0 \rangle|^2 = \hbar^2/2$.
In Eq.~(\ref{IntDecayMG1}) $k$ stands for $k(n)$.
Equation~(\ref{IntDecayMG1}) is written using the dimensionless parameters $q$ and $\eta$ as
\begin{eqnarray}
	\Gamma_n &=& \wosc\frac{\oldalpha^2}{2^{n+2} n! q \sqrt{\pi}}
	 \times \left| \int_{ -\infty}^{\infty}\!\!\! du H_n(u) e^{-u^2/2} \left[ \frac{u \left(e^{i qu} + (-1)^{n+ 1} e^{-i qu}\right)}{(u^2+\oldalpha^2)^2} - i \frac{q \left(e^{i qu} + (-1)^{n} e^{-i qu}\right)}{u^2+\oldalpha^2}\right] \right|^2 \label{IntDecayMG}
\end{eqnarray}
\end{widetext}
where $q$ stands for $q(n)$, as given in Eq.~(\ref{eq:q.def}).
Numerical results obtained from Eq.~\eqref{IntDecayMG} are presented
on Fig.~\ref{fig:fig2} (and also on Fig.~\ref{fig:LZOmega} which will
be discussed in the next section~\ref{sec:low-gradient-model}).

An analytical solution for the ground state case with $n=0$ can be
found (where $ H_0(u) = 1$). To accomplish this we use the integrals
3.954.1 and 3.954.2 from Gradshteyn and Ryzhik \cite{GradshteynRyzhik}
to find
\begin{eqnarray}
	\Gamma_0 &=& \wosc\frac{\pi^{\frac{3}{2}} }{16 q_0} e^{\oldalpha^2} \nonumber
	\\ &&  \times \Bigg\{ e^{- \oldalpha q_0} \left( q_0 + \oldalpha
        \right) \mathrm{erfc} \left[ \frac{-1}{\sqrt{2}} \left( q_0 -
            \oldalpha \right) \right]  \nonumber
	\\ && + e^{\oldalpha q_0} \left( q_0 - \oldalpha \right)
        \mathrm{erfc} \left[ \frac{1}{\sqrt{2}} \left( q_0 + \oldalpha  \right) \right] \Bigg\}^2, \label{transboxho}
\end{eqnarray}
which is expressed in terms of `erfc' the complementary error function
\cite{AbramowitzStegun}, the dimensionless variable $\oldalpha$ and
the scaled dimensionless momentum $q_0=k_{(n=0)}
\oldsigma=\sqrt{2\oldalpha^2+1}$. We see that $q_0$ depends on
$\oldalpha$ and that $q_0>\oldalpha$.

For higher energy trapped atoms with $n \geq 1$ the integrals
contained within Eq.~\eqref{IntDecayMG} can be approximated in the
region $\eta\gtrsim 5$ by calculating the residue of a pole found
within them. This leads to an analytic expression for the decay rates
provided by Fermi's Golden Rule for any $n$ state \cite{BurrowsThese}:
\begin{eqnarray}
	\Gamma_n &\approx&\wosc \frac{\pi^{\frac{3}{2}} }{2^{n+2} n! q }
        \times \exp \left( {\oldalpha^2 - 2 \oldalpha q} \right) \nonumber
	\\ && \times \left| 2 n H_{n-1}\left(i \oldalpha \right)  - i \left( q + \oldalpha \right) H_n \left(i \oldalpha \right)\right|^2,   \label{Gamman}
\end{eqnarray}
where, again, $q$ stands for $q(n)=k(n)\oldsigma$,
Eq.~(\ref{eq:q.def}), which does depend on $\oldalpha$.  Results from
this expression are shown for a specific example in
Fig.~\ref{fig:fig2} where good agreement is seen with the numerical
evaluation of Eq.~\eqref{IntDecayMG} (solid line in that figure).

It is useful to find the $\Rzero \rightarrow \infty$ and
$B^\prime\rightarrow 0$ (or equivalently $\alpha\rightarrow 0$) limits
of Eq.~\eqref{Gamman}, as it is in these regimes that cold atom traps
favourably operate as the losses due to non-adiabatic effects are low.
Since $\Gamma_n/\wosc$ depends only on $\oldalpha$ and $n$ in
Eq.~(\ref{Gamman}), and since
$\oldalpha\propto\Rzero^{3/4}/\alpha^{1/2}$, see Eq.~(\ref{eq:eta}),
both limits are found from $\oldalpha\rightarrow\infty$.  The limiting
behaviour, valid for states $n$ such that $n\ll \eta^2$, is then
\begin{equation}
	\Gamma_n/\wosc  \underset{\oldalpha\rightarrow\infty}{\sim} 
        \frac{2^n}{n!}
        \oldalpha^{2n+1} e^{-\sqrt{2}(n+\frac{1}{2})} e^{ - (2\sqrt{2}-1 )\oldalpha^2 }.
\label{Hi-eta-limit}
\end{equation}
We will see in Sec.~\ref{sec:comp-with-LZ} that this behaviour agrees
qualitatively with our semi-classical interpretation
(Sec.~\ref{sec:landau-zener-theory}).  The trap frequency
\eqref{omega} is increased for high magnetic field gradient or low
Rabi frequency which raises the energy of the $n$th oscillator level
and increases crossing speed.  Additionally, tighter trapping
potentials lead to the orientation of the local effective magnetic
field direction changing more rapidly over a given distance. Both of
these factors result in a greater probability for an atom to become
misaligned from the local effective magnetic field vector, leading to
greater non-adiabatic losses as $\Rzero \rightarrow 0$ or $B^\prime
\rightarrow \infty$.  Equation~(\ref{Hi-eta-limit}) is useful for
specifying the main dependence of non-adiabatic decay rates on
$\Rzero$ and $B^\prime$, additionally indicating that the process is
more sensitive to Rabi frequency than magnetic field gradient since
$\oldalpha^2 \propto \Rzero^{3/2}/ B'$.

\subsection{\FG\ model}
\label{sec:low-gradient-model}

In the case where the $z$ axis is oriented vertically we can no longer
neglect the effect of gravity on the location of the equilibrium
position of the atoms in the adiabatic potential. This time the
harmonic expansion yields an initial trapping potential
\begin{equation}
	V_i (z) = V_0 + \frac{1}{2} \mass {\wosc}^2 (z-z_0)^2, \label{HOgravity} 
\end{equation}
where, because of gravity, the centre of the harmonic oscillator is shifted
from the origin (where there is resonance) to a point
\begin{equation}
z_0 = -
\frac{\epsilon \Omega_0}{\alpha\sqrt{1 - \epsilon^2}}
\label{eq:z0}
\end{equation}
below the origin. The parameter $ \epsilon = \mass g / \hbar \alpha $
is introduced as the ratio of the gravitational force to the force
applied by the magnetic field gradient. The approximate harmonic
potential now has a modified energy offset (compared to
Eq.~(\ref{HOMG}))
\begin{equation}
  V_0 = \hbar \Rzero \sqrt{1 - \epsilon^2}
 \label{eq:V-offset}
\end{equation}
and a modified trap frequency \cite{Merloti2013a}
\begin{equation}
	\wosc = \alpha \sqrt{\frac{\hbar }{\mass \Rzero}} 
               {\left( 1 - \epsilon^2 \right)}^{\frac{3}{4}}. \label{omegag}
\end{equation}
We see that if $\epsilon=0$ we recover the \mg\ model result
Eq.~(\ref{omega}), and also that there is no trap unless
$\epsilon<1$. This condition is equivalent to gravity compensation by
the magnetic force in the underlying static magnetic trap, namely
$\hbar\alpha>\mass g$.

The wavefunction for an atom in the initial trapped $\mfp = 1$ state
is now a \emph{displaced} harmonic oscillator wavefunction,
\begin{equation}
        \psii = \frac{H_{n}[(z-z_0)/\oldsigma]
         }{\sqrt{n! 2^{n}\oldsigma\sqrt{\pi}}} \, e^{{-(z-z_0)^2/(2 \oldsigma^2)}}\label{GravityHO}
\end{equation}
where $ \oldsigma = \sqrt{\hbar/\mass \wosc} $ should be used with the
appropriate trap frequency $\wosc$, Eq.~(\ref{omegag}), and $n$ is a
positive integer which selects the energy of the trapped atom from the
allowed discrete harmonic oscillator energy levels given by the
relevant $ E_{n} = \left( n + \frac{1}{2} \right) \hbar \wosc + V_0$.

\begin{figure*}[t]
\includegraphics[width=0.80\linewidth]{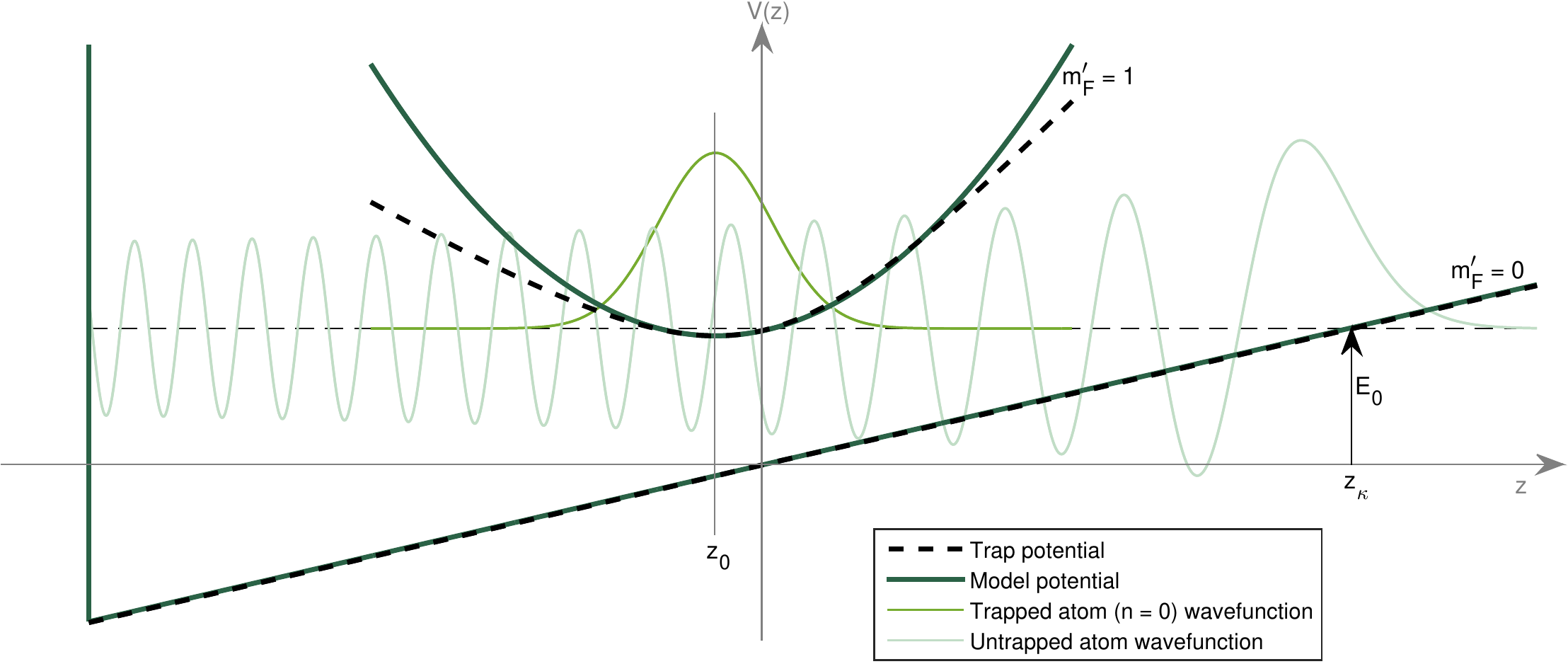}
\caption[\FG\ model potentials]{Schematic diagram showing the key
  wavefunctions and potentials $V_{i,f}(z)$ in the \fg\ model. The
  wave-functions are scaled to equal maximum height for comparison.
  The ground state wavefunction of the harmonic oscillator is shown
  centred at $z_0$, Eq.~(\ref{eq:z0}), in its potential $V_i(z)$,
  Eq.~(\ref{HOgravity}). The energy of the ground state is $E_0$, as
  given by $V_0 + \hbar\wosc/2$,
  Eqs.~(\ref{eq:V-offset},\ref{omegag}).  An energy resonant Airy
  function is shown, which is an eigenstate of the linear potential
  $V_f(z)$, Eq.~(\ref{Vslope}).  The potential `wall' on the left is
  located at $z\rightarrow -\infty$ in the calculations. The
  eigenstate is associated with a turning point at $z_\kappa$. The
  parameters for this figure are as in
  Fig.~\ref{fig:EZRF}.\label{fig:fig3}}
\end{figure*}

The potential for an untrapped atom ($\mfp = 0$) is now simply:
\begin{equation}
 V_f(z) = \left\{
  \begin{array}{l l}
     \infty, & \quad z \leq -\oldGravityBoxL, \\
      \mass g z, & \quad z > -\oldGravityBoxL. \\
  \end{array} \right.   \label{Vslope}
\end{equation}
To assist with the calculation we have introduced a distance
$\oldGravityBoxL$ to a (single) hard
wall of the potential, see Fig.~\ref{fig:fig3}, similarly to the distance $L$ in
Sec.~\ref{sec:high-gradient-model}. Later we will also let $\oldGravityBoxL\rightarrow\infty$.
The corresponding stationary Schr\"odinger equation in the final state $\mfp = 0$ is given by
\begin{equation}
	E_\kappa \psifa = - \frac{\hbar^2}{2 \mass} \frac{\mathrm{d}^2
          \psifa}{\mathrm{d} z^2} + \mass g z \psifa
\label{SEfg}
\end{equation}
where $\kappa$ is an index labelling the final state.  Equation
\eqref{SEfg} can be written in terms of a second order partial
differential equation for a spatially shifted Airy function, $
\frac{\mathrm{d}^2}{\mathrm{d} \zeta^2} \mathrm{Ai}(\zeta)= \zeta
\mathrm{Ai}(\zeta)$ where the argument $\zeta$ is given by
$\zeta=\tilde z -\tilde z_\kappa $, and we use the scaled distance $
\tilde z = z / \ell$.  The characteristic length of the Airy function
is linked to the atom mass and gravity through
\begin{equation}
\ell =  \left( \frac{\hbar^2}{2 \mass^2 g} \right)^{1/3}
\label{eq:def-beta}
\end{equation}
and the turning point for a classical particle is at $ z_\kappa =
E_\kappa/ \mass g = \tilde z_\kappa \ell $.  Fermi's Golden rule will
require energy matching of the $\mfp = 0$ and $\mfp = 1$ states as
seen for $n=0$ in Fig.~\ref{fig:fig3}. In terms of $z$, the solution
of the stationary Schr\"odinger equation for the untrapped state is
then \cite{AbramowitzStegun,Gerbier2001}
\begin{equation}
  \psifa = \AiryConstant \mathrm{Ai} \left[\left( z - z_\kappa \right)
    /\ell \right] 
 \,,
\label{GravitySlope}
\end{equation}
where $\AiryConstant$ is a normalisation constant and $\mathrm{Ai}$ is
the Airy function of the first kind.

The normalisation constant $\AiryConstant$ for the untrapped state
wavefunction can be determined from the condition that $
{|\AiryConstant|}^2 \int_{-\oldGravityBoxL}^{\infty} {|\mathrm{Ai}
  \left[ \left( z - z_\kappa \right) /\ell \right]|}^2 dz=1$. If the
Airy function $ \mathrm{Ai}(\zeta)$ is approximated in the
$\zeta\rightarrow -\infty$ limit by \cite{AbramowitzStegun} $
\mathrm{Ai}(\zeta) \approx \frac{1}{\sqrt{\pi} {\left( - \zeta
    \right)}^{\frac{1}{4}}} \sin{\left[ \frac{2}{3} {\left( - \zeta
      \right)}^{\frac{3}{2}} + \frac{\pi}{4} \right]} $, this leads to
an approximate normalisation constant for the untrapped state
wavefunction given by ${|\AiryConstant|}^2 \approx \pi / \sqrt{ \ell
  \oldGravityBoxL}$.  In the continuum limit, and to apply Fermi's
Golden Rule, we need the density of states. To determine this we note
that the potential wall at $z=-\oldGravityBoxL$ creates a boundary
condition for the wavefunction as $ \psifBC = 0$. The asymptotic form
of the Airy function can be used to express this condition as a
quantization condition since the argument of the sine function should
be a multiple of $\pi$. Specifying the multiple by the integer
$n_\kappa$, we have the condition $ n_\kappa \pi = \frac{2}{3} {\left[
    \left(z_\kappa + \oldGravityBoxL\right)/\ell
  \right]}^{\frac{3}{2}} + \frac{\pi}{4}$.  Differentiation of this
quantization condition, together with $E_\kappa=\mass g z_\kappa$,
leads to an equation for the density of states,
\begin{eqnarray}
	D \left( E_\kappa\right ) &=& \frac{\partial n_\kappa}{\partial E_\kappa} =
        \frac{1}{\pi (\mass g \ell)^{3/2}}  \sqrt{  E_\kappa + \mass
          g \oldGravityBoxL  }
          \nonumber\\
        &\underset{\oldGravityBoxL\rightarrow\infty}{\simeq} &
        \frac{1}{ \pi \mass g \ell}\sqrt{\frac{\oldGravityBoxL}{\ell}}.
\end{eqnarray}
It is noteworthy that the $\oldGravityBoxL$ dependence cancels in the
product ${|\AiryConstant|}^2 \cdot D \left( E_\kappa\right ) =1/(\mass
g \ell^2) $ such that there are no issues when taking the
$\oldGravityBoxL \rightarrow \infty$ limit.

The interaction matrix element associated with the non-adiabatic coupling is given by
\begin{equation}
         \frac{ \sqrt{2} \hbar}{2 i} \int_{ - \infty}^{\infty}
         \psiicc \hat{V}_A \psifa
dz  
\label{integral-overlap-gravity}
\end{equation}
where, as in the \mg\ model, the factor $2i$ arises from the component
of $\hat F_-$ in $\hat F_y$ (see Eq.~\eqref{Hamiltonian} and the
factor $\sqrt{2}\hbar$ comes from the matrix element of $\hat F_-$
between $\mfp=1$ and $\mfp=0$. In evaluating this integral the
wavefunctions will be given by equations \eqref{GravityHO} and
\eqref{GravitySlope}.

\begin{figure}[t]
\includegraphics[width=\linewidth]{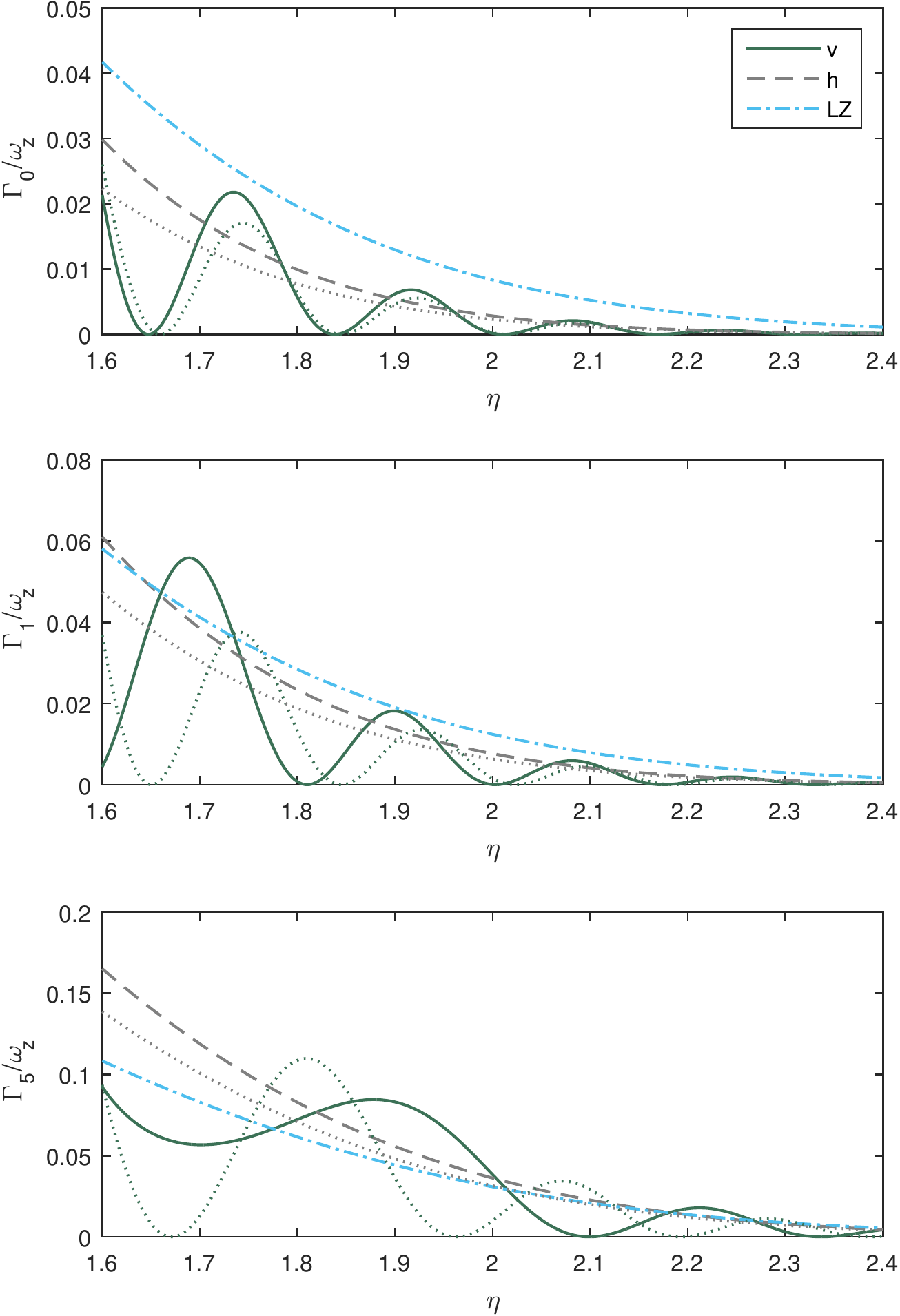}
\caption[Decay rate variation with $\oldalpha$]{Scaled decay rates
  $\Gamma_n/\wosc$ as a function of the parameter $\oldalpha$ for the
  ground state ($\Gamma_0/\wosc$, top), the first excited state
  ($\Gamma_1/\wosc$, middle) and the fifth excited state
  ($\Gamma_5/\wosc$, bottom) of the initial harmonic trap. The
  prediction of the \fg\ model, Eq.~\eqref{IntDecayFG} solid line, is
  shown with the predictions of the \mg\ model, Eq.~\eqref{IntDecayMG}
  dashed line, and the Landau-Zener model, Eq.~\eqref{LZDecay} chained
  line.  The dotted lines indicate the effect of non-adiabatic
  potentials which produce corrections to $\wosc$ from
  Eq.~\eqref{eq:AppendixFinal-wosc} and to $\Gamma_n$ as described in
  Appendix~\ref{sec:fg-model-general} and in Fig.~\ref{fig:fig2}. The
  calculations are done for the $F=1$ hyperfine ground state of
  $^{87}$Rb with $\epsilon\sim 0.20$ in the \fg\ model case.
\label{fig:LZOmega}}
\end{figure}

All the components necessary for use of Fermi's Golden Rule are now
known, and putting this together we find the decay rate for the $n$th
oscillator state in the \fg\ model
\begin{eqnarray}
	\Gamma_{n} &=& \wosc \frac{2 \sqrt{\pi} \oldalpha^2 }{ n! 2^{n} \beta} \nonumber
	\\ &\times&  \bigg| 
	\int_{ - \infty}^{\infty} \frac{ \left(
            u + u_0 \right) H_{n}\left(u \right) e^{-\frac{u^2}{2}}}{{\left[ (u+u_0)^2 + {\oldalpha}^2 \right]}^2} \mathrm{Ai}\left[\beta(u+u_0-u_\kappa) \right] d u \nonumber
	\\ &-&   \beta \int_{ - \infty}^{\infty} 
          \frac{ H_{n}\left(u\right)
          e^{-\frac{u^2}{2}}}{ \left( u + u_0 \right)^2 + \oldalpha^2}
           \mathrm{Ai}' \left[ \beta(u+u_0-u_\kappa) \right] d u 
            \bigg|^2 \label{IntDecayFG}
\end{eqnarray}
where we integrate over $u = (z-z_0)/\oldsigma$ and we have defined
$u_0 = z_0/\oldsigma$, $u_\kappa= z_\kappa/\oldsigma = E_\kappa/
(\mass g \oldsigma )$, $E_\kappa = E_n$, $\beta=\oldsigma/\ell$ and
$\oldalpha=\Omega_0/(\alpha\oldsigma)$. (Note that $\eta$, $\wosc$,
and $\oldsigma$ differ from the expressions of
Sec.~\ref{sec:high-gradient-model}.)  The function $\mathrm{Ai}'$ is
the usual derivative of the Airy function with respect to its
argument.  This result for $\Gamma_{n}/\wosc$ can be expressed solely
in terms of $n$, $\oldalpha$ and $\epsilon$ since the integrals in
(\ref{IntDecayFG}) depend only on $n$, $\oldalpha$, $\beta$, $u_0$ and
$u_\kappa$ and with the above definitions it can be shown that
\begin{eqnarray}
\beta^3 &=& (\oldsigma/\ell)^3 = 2 \oldalpha \epsilon ( 1 - \epsilon^2 )^{-3/2} 
\,,  \nonumber\\
u_0 &=& z_0/\oldsigma = - \oldalpha \epsilon ( 1 - \epsilon^2 )^{-1/2} 
\,,\nonumber\\
u_\kappa &=&  \frac{ V_0 + (n+1/2)\hbar \wosc }{\mass g \oldsigma} \nonumber\\
&=& 
   \frac{\oldalpha}{\epsilon} ( 1 - \epsilon^2 )^{1/2} 
   \left[
 1 +  \frac{ 1 - \epsilon^2 }{  \oldalpha^2  }\left( n + \frac{1}{2}\right)
   \right]
\,.   
\label{eq:beta-u0-uk-in-terms-of-eta-and-epsilon}
\end{eqnarray}
Gravitational effects are very weak if $\epsilon$ is small
($\epsilon\ll 1$) and we see in Sec.~\ref{sec:comp-with-LZ} that the
adiabatic limit is reached if $\oldalpha$ is large ($\oldalpha\gg 1$).

Figures \ref{fig:fig2} and \ref{fig:LZOmega} show numerical results
obtained from Eq.~\eqref{IntDecayFG} for the scaled decay rate
$\Gamma_{n}/\wosc$ as a function of the initial quantum number $n$ in
the harmonic trap and of the parameter $\oldalpha$ given in the
general case by
\begin{equation}
\oldalpha = \left(\frac{\mass \Omega_0^3}{\hbar\alpha^2}\right)^{1/4}\left(1-\epsilon^2\right)^{3/8}.
\label{eq:def_eta}
\end{equation}
We note that we have approximately $\oldalpha\propto
\Omega_0^{3/4}B'^{-1/2}$ when $\epsilon$ stays small (as in the \mg\
case). The parameter $\oldalpha$ thus remains more sensitive to the
Rabi frequency, proportional to the amplitude of the dressing field,
than to the magnetic gradient.

In contrast to the predictions of the \mg\ model, the \fg\ model
displays in Fig.~\ref{fig:LZOmega} a clear oscillatory behaviour in
the decay rate variation with $\oldalpha$, or equivalently when
varying Rabi frequency or magnetic field gradient. Additionally, in
contrast to the semi-classical interpretation of non-adiabatic losses,
there is not a monotonic increase of the decay rate with atomic energy
or vibrational level $n$, with some high $n$ states being sheltered
from non-adiabatic losses.

\begin{figure}[t]
\includegraphics[width=\linewidth]{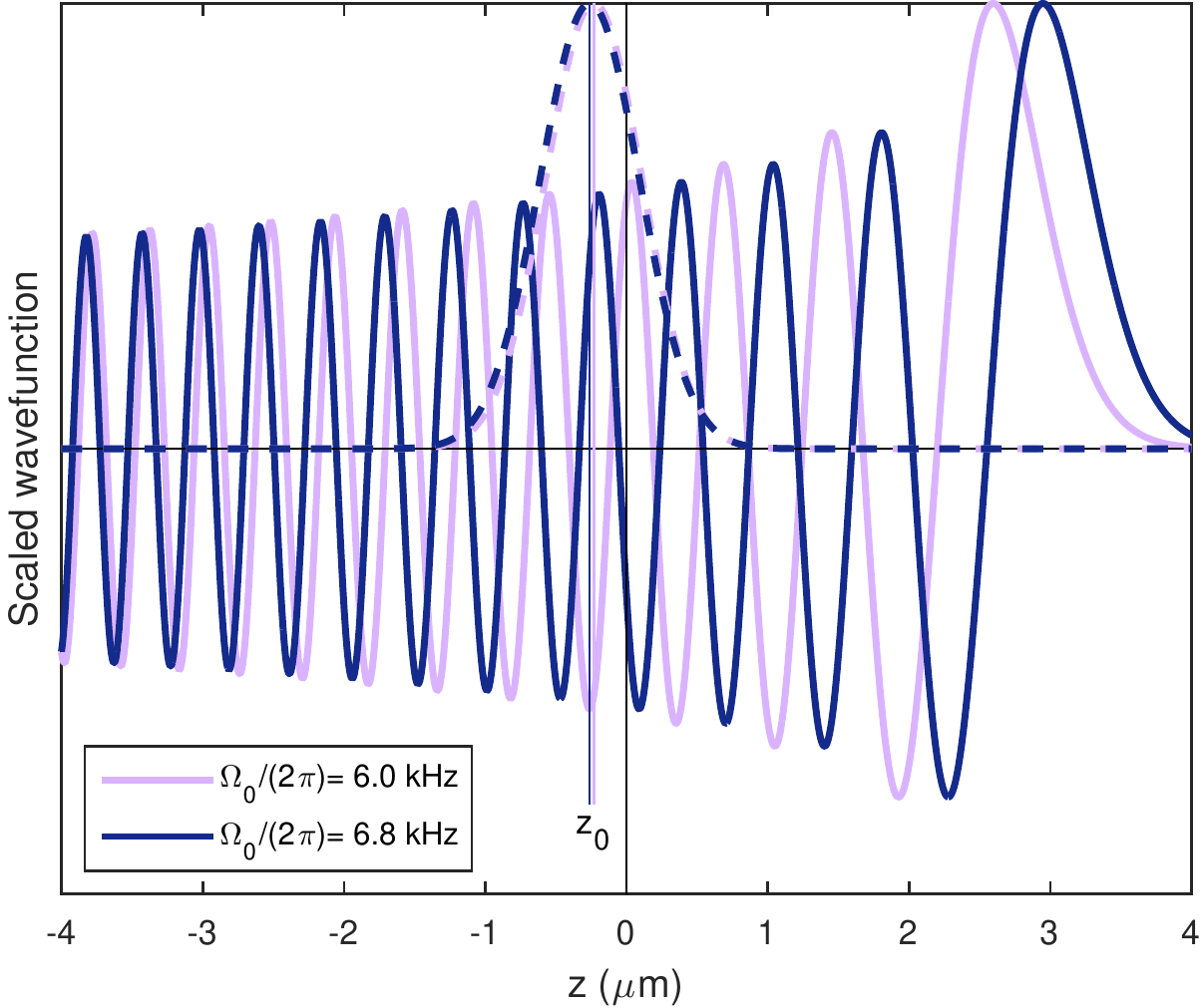}
\caption[\FG\ model wavefunctions varying with Rabi frequency]{An
  examination of the sensitivity of the wavefunctions in the \fg\
  model. The trapped harmonic oscillator ground state wave-function
  $\psii$, associated with $\mfp = 1$, is displayed with dashed lines
  for two different Rabi frequencies: $\Rzero/(2\pi)= 6.0$~kHz and
  $6.8$~kHz. Other parameters are fixed and are the same as in
  Fig.~\ref{fig:EZRF}.  Also shown are the two Airy functions
  associated with $\mfp=0$ eigenstates matching the energies of the
  corresponding harmonic oscillator ground states. We see that there
  is a substantial shift in the peaks of the Airy functions between
  the two Rabi frequencies shown, whilst the change in the location of
  the oscillator ground state, $z_0$, is very small. This shows how
  the overlap integral Eq.~(\ref{integral-overlap-gravity}) can be
  sensitive to parameters.  The wavefunctions have been scaled so that
  they reach a value of unity at the maximum
  height.\label{fig:GravityWaveFcns}}
\end{figure}

To help understand the origin of the oscillatory behaviour of the
decay rates, Fig.~\ref{fig:GravityWaveFcns} shows the shape of the
trapped and untrapped state wavefunctions for two different Rabi
frequencies.  The vertical lines marked $z_0$ are nearly coincident in
Fig.~\ref{fig:GravityWaveFcns} indicating that the centre of the
harmonic oscillator hardly shifts when the Rabi frequency is changed.
However, the energy of the untrapped state is set to match that of the
trapped atom as needed to satisfy Fermi's Golden Rule.  This means
that changing the harmonic oscillator energy level structure (for
example, by changing the Rabi frequency) affects the Airy
wavefunction.  Thus, when the Rabi frequency is changed, we see that
although the displacement in the minimum $z_0$ is weakly affected, the
oscillations of the relevant Airy function are significantly
displaced.  It is this progression of the Airy function peaks which
leads to the oscillatory behaviour in the decay rates obtained as a
function of $\oldalpha$ (and $\epsilon$) and seen in
Fig.~\ref{fig:LZOmega}. (This oscillatory dependence on $\oldalpha$
can also be found equivalently as a function of the unscaled
parameters $\Rzero$ or $B'$.)  In other words, for the case of
horizontal trapping, the phase of a plane wave can always be set to
match the location of the atom, but in the case of vertical trapping,
the oscillatory phase of the Airy function is restricted by the energy
of the initial state.  For vertical trapping, the behaviour of
$\Gamma_n$ as a function of $n$ is not smooth as there is a dramatic
change in the harmonic oscillator wavefunction $(\Phi_n)$ with the
quantum number $n$, which affects the interaction matrix element and
leads to the results seen in Fig.~\ref{fig:fig2}.

\section{Comparison of quantum dynamics with the Landau-Zener model}
\label{sec:comp-with-LZ}

In this section we compare the quantum decay rates obtained in
Sec.~\ref{sec:quantum-dynamics-1d-trap} to the Landau-Zener model
introduced in Sec.~\ref{sec:landau-zener-theory}.  Returning to
Eq.~\eqref{LZgeneral}, we note that the concept of atomic speed relies
on the idea of a classical trajectory.  We stress here that the
classical trajectory in the bare (uncoupled) potential should be used
to compute the transition rate \cite{Garraway1991}.  Although the
trajectory is described classically, for a more direct comparison with
our quantum mechanical decay rates it is beneficial to describe the
Landau-Zener decay rate in terms of the atomic energy level denoted by
the quantum number $n$. By considering energy conservation of an atom
at the resonance location, the expression $ \frac{1}{2} \mass v^2 =
E_n = V_0+\hbar \wosc \left( n + \frac{1}{2} \right) $ is obtained,
which leads to
\begin{equation}
  \label{eq:v1}
v = \sqrt{\frac{2V_0}{\mass} + (2n+1)\,\wosc^2\oldsigma^2}
\end{equation}
for the atomic speed through the resonance location. Therefore the
Landau-Zener decay rate (\ref{eq:LZ-adiabatic-limit}) for an $F=1$
atom in the $n$th harmonic oscillator energy level is
\begin{widetext}
\begin{equation}
	\Gamma_n^{LZ}
 = \frac{\wosc}{\pi} \left\{ 1 - {\left[  1 - \exp \left( - \frac{\pi
           \oldalpha^2}{2\sqrt{2}(1-\epsilon^2) \sqrt{1+\frac{1}{\oldalpha^2}(n + \frac{1}{2})(1-\epsilon^2)}} \right) \right]}^{2} \right\}. 
\label{LZDecay}
\end{equation} 
In the limit $\epsilon\rightarrow 0$ we obtain the result for
horizontal trapping, that is
\begin{equation}
	\Gamma_n^{LZ}
 = \frac{\wosc}{\pi} \left\{ 1 - {\left[  1 - \exp \left( - \frac{\pi
           \oldalpha^2}{2\sqrt{2} \sqrt{1+\frac{1}{\oldalpha^2}(n + \frac{1}{2})}} \right) \right]}^{2} \right\}.
\label{LZDecayMG}
\end{equation}
\end{widetext}
To simplify the expression further we note that for the lowest
harmonic levels, where the harmonic approximation for the adiabatic
potential is valid ($n\ll \oldalpha^2$ and $\eta\gg1$), we obtain
\begin{equation}
	\Gamma_n^{LZ}
\simeq \frac{2\wosc}{\pi}  e^{(2n+1)\pi/(8\sqrt{2})}
\exp\left(-\frac{\pi \oldalpha^2}{2\sqrt{2}} \right). 
\label{LZDecaylimit}
\end{equation} 
Comparison of Eq.~(\ref{Hi-eta-limit}) with the Landau-Zener model
limiting behaviour given in Eq.~\eqref{LZDecaylimit} shows some
structural similarity, but also clear differences between our model
and the Landau-Zener model in the low decay regime.  The similarity is
the exponential dependence on $\oldalpha^2$ with a slight difference
in the multiplying factors, i.e.\ a factor of $\pi/(2\sqrt{2})\simeq
1.1$ in the Landau Zener case, and a factor of approximately $1.8$ in
the case of Eq.~(\ref{Hi-eta-limit}). For large $\oldalpha$, the
desired limit for trap operation, this exponential dependence is the
most dominating aspect and leads to an overestimation of the decay
rate by the Landau-Zener model.  Another difference between the
results, which is more relevant at lower $\oldalpha$, is that the
polynomial pre-factor, with its power-law dependence
$\oldalpha^{2n+1}$, is absent in the Landau-Zener model.

The overestimation of the Landau-Zener model is seen in
Fig.~\ref{fig:fig2}. In particular it is most clearly seen for the
vibrational ground state ($n=0$), and lower $n$ values, which are
often dominantly populated at the low temperatures necessary for
ultra-cold atom traps.  The Landau-Zener result improves in comparison
with the higher $n$ values of the horizontal model in
Fig.~\ref{fig:fig2}, however, when comparing it to the vertical model
there are the irregular oscillations in the decay rate as discussed in
Sec.~\ref{sec:low-gradient-model}. These kinds of oscillations can not
be obtained from the simple application of the Landau-Zener model to a
single crossing. However, although the Landau-Zener model generally
overestimates the decay rate, there are a few points, at higher $n$,
where the oscillatory \fg\ model decay rate slightly exceeds the
Landau-Zener result.

\begin{figure}[t]
\includegraphics[width=\linewidth]{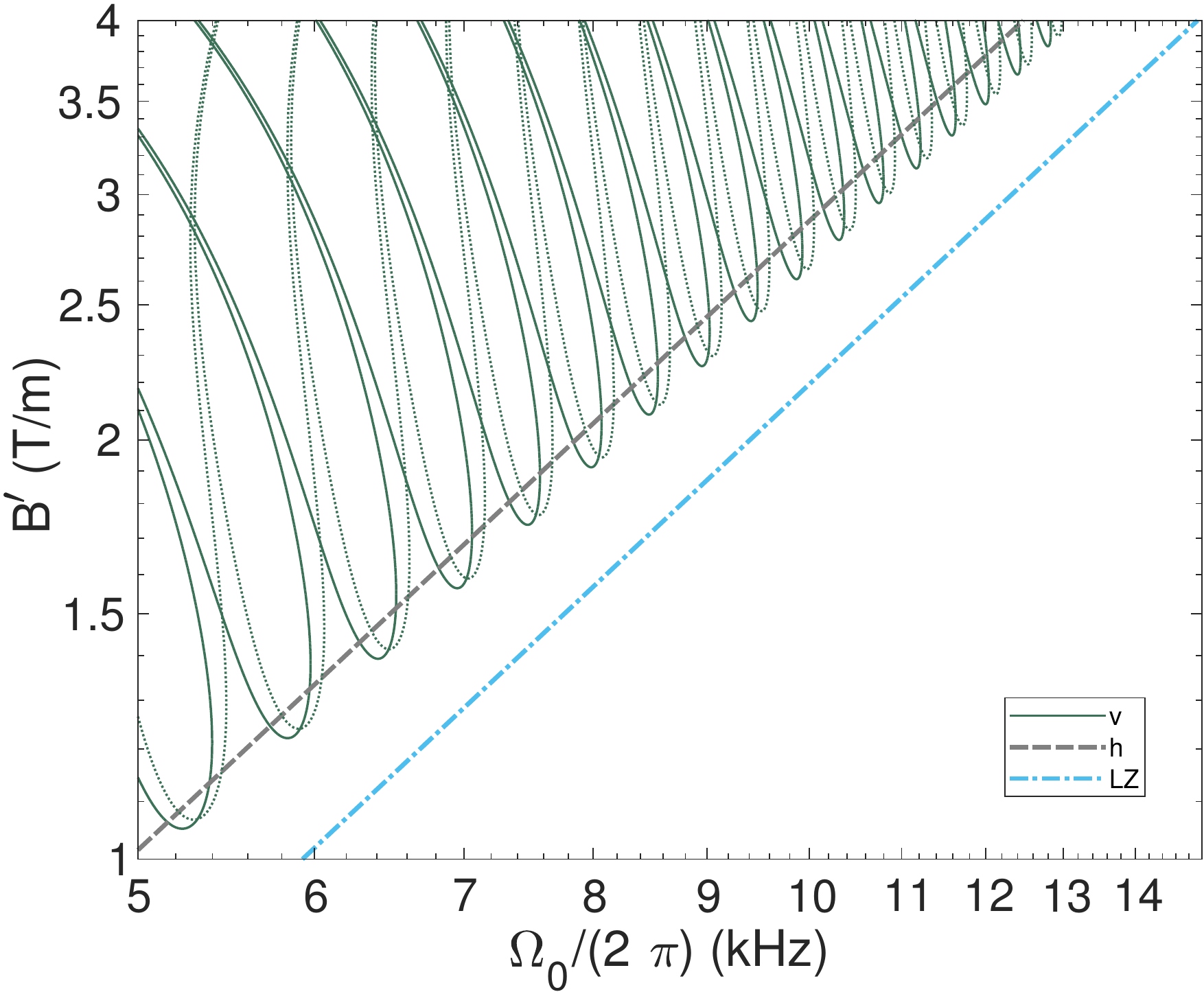}
\caption[]{Ground state decay rate contours for different models as a
  function of Rabi frequency and field gradient.  The contours show
  where the scaled decay rate $\Gamma_0/\wosc = 10^{-3}$, i.e.\ where
  decay takes place after about 160 oscillations in the trap.  The
  decay rates generally increase towards the top left of the figure.
  The solid contour line (v) shows the result from the \fg\ model, as
  given by Eq.~\eqref{IntDecayFG}.  The dotted contour line shows the
  effect of including the non-adiabatic potentials from $\hat V_B$ in
  the \fg\ model case (see Appendix~\ref{sec:fg-model-general}).  The
  dashed contour line corresponds to the ground state decay rate from
  the \mg\ model (h), as given by Eq.~\eqref{IntDecayMG}. It is a
  suitable approximation to the boundary displayed by the \fg\ model.
  The effect of including the non-adiabatic potentials is not shown
  for the \mg\ model as there is no visible difference from the dashed
  contour line.  The chained contour line corresponds to the
  Landau-Zener decay rate, Eq.~\eqref{LZDecay}, with a trap frequency
  calculated using Eq.~\eqref{omegag}.\label{fig:LZOmegaBGrad} The
  results presented in this figure are for ${}^{87}$Rb and $F=1$.}
\end{figure}

Figure \ref{fig:LZOmegaBGrad} shows how the Fermi Golden Rule decay
rates imply less stringent requirements on trap Rabi frequency and
magnetic field gradient in comparison to the Landau-Zener model
prediction for a given ground state lifetime. Here we use the direct
experimental parameters $\Rzero$ and $B'$ to clearly indicate the
practical consequences of the results for ${}^{87}$Rb.  We see that
the Landau-Zener model provides useful guidance: given the logarithmic
scales of the figure, the power-law dependence of Landau-Zener result
on the parameters is approximately correct, with a consistent
margin. When adiabaticity is reduced below the Landau-Zener boundary
in Fig.~\ref{fig:LZOmegaBGrad} the oscillatory structures of the \fg\
model appear. The details of these are sensitive to phase shifts from
non-adiabatic corrections to the potentials, as seen from the dotted
lines.

\section{Conclusion}
\label{sec:conclusion}

Since their experimental inception in 2004 \cite{Colombe2004a},
rf-dressed adiabatic potentials for cold atoms have been successfully
applied to a wide variety of situations \cite{Garraway2016}. Generally
speaking, for the trap to work effectively, one is content to ensure
that a sufficiently strong rf field is employed so that as few atoms
as possible are lost from the trap. However, if an over-strong rf
field is used the trapping frequency of the trap itself is reduced,
which is undesirable if low-dimensionality is required
\cite{QGLD2003}. Furthermore, an over-strong rf field can result in
infringement of the RWA \cite{Hofferberth2007}. And there may be other
situations, for example involving time-averaged adiabatic potentials
or multi-frequency adiabatic potentials
\cite{Garraway2016,Perrin2017}, where the Rabi frequency is required
to be constrained. Then the design question arises as to how small the
coupling can be made before the trap is no longer effective.

In this paper we have tried to address this complex situation with a
number of significant approximations, but also with the aim of
obtaining some analytic results. We have had to treat a
one-dimensional quantum problem as the three-dimensional problem is
not separable due to the changing direction of gravity and the typical
change in relative orientation of the magnetic and rf fields around
the full 3D trapping surface \cite{Perrin2017}. The approximation
seems reasonable for those cases where the atom cloud is sufficiently
compact. We have assumed that the rf polarisation is linear, and in a
specific direction. However, in this case the results can be
generalised quite easily to the case of other polarisations. We have
also assumed that the atomic Rabi frequency is uniform for the spatial
$z$ variation that we consider. As discussed, this is a reasonable
approximation for some situations (macroscopic coils for rf radiation),
but needs more careful consideration in other cases such as when atom
chips are being used.  We have also assumed the linear Zeeman effect
(though for a treatment of adiabatic potentials in the non-linear
regime see Ref.~\cite{SinucoLeon2012}). The final expressions for
decay rates have assumed a local harmonic approximation for the
adiabatic potential. This assumption places a constraint on the
excitation in the trapping degree of freedom (i.e.\ $n \ll
\oldalpha^2$). Of course, this constraint could be relaxed in a fully
numerical approach to the problem where the harmonic oscillator basis
states are not used in the evaluation of the matrix elements.

Fermi's Golden rule (i.e.\ time-dependent perturbation theory) has
been a key tool to determine the decay rates. This relies on a weak
coupling between the initial and final states of the model and is
expected to be very appropriate in the adiabatic limit. However, in
assuming that the coupling is weak we have neglected the possibility
of non-Markovian dynamics during the decay process. This is reasonable
as such dynamics would only be expected to appear when the atom loss
is very rapid and it would be difficult to observe.  We have also
focused on the case $F=1$ in this paper. Aspects of the derivation can
be easily generalised to higher $F$, however, it may not be possible
to use Fermi's Golden rule any more because the $\hat V_A$ couplings
are no longer to a continuum when the uppermost adiabatic state
($\mfp=F$, $F>1$) is considered for non-adiabatic loss. However, the
good news is that it is likely that because the transitions are now
bound-bound rather than bound-continuum the loss of atoms from the
upper-most adiabatic state is expected to be strongly inhibited; that
is, you would be `unlucky' to find a coincidence of the vibrational
eigen-energies of $\mfp=F$ and $\mfp=F-1$ because the harmonic
frequencies are not commensurate with each other and are also shifted
by an incommensurate Rabi frequency.  We remark however that the role
of the transverse ($x$ and $y$) directions, which we do not take into
account here, could compensate for this energy mismatch and restore
the losses.  An additional approximation in the work presented here is
the neglect of that part of the $\hat V_B$ coupling that causes a
downward change in $\mfp$ by two. In the case of $F=1$ this would add
the complication of an additional density of final states on an
inverted ($\mfp=-1$) potential as well as possible second order $\hat
V_A$ processes that could coherently interfere with it. Numerical work
has suggested that these effects can be neglected for the parameters
we have considered here, but the effects could become significant in
other situations and it would be good to quantify this in the
future. However, the part of the $\hat V_B$ coupling that causes
non-adiabatic potentials to be added to the $\mfp$ states has been
included. These are based on the analysis in the
Appendix~\ref{sec:general-spin-formulae} and included as dotted lines
for $\Gamma_n$ in Figs.~\ref{fig:fig2}, \ref{fig:LZOmega}
and~\ref{fig:LZOmegaBGrad}.  Finally, in this recapitulation of the
approximations, we have not included consideration here of some very
practical matters such as the losses due to collisions with background
gas atoms and molecules, or the effects of heating due to noise in the
currents that may produce either the static magnetic field or the rf
magnetic field.

Despite these approximations, the results for horizontal trapping,
Eq.~\eqref{Gamman}, and vertical trapping, Eq.~\eqref{IntDecayFG},
should be able to offer some safe guidance with an appropriate
estimate for the excitation $n$, or with a distribution of $n$ such as
can be found with a thermal state.  To obtain the results we used both
Landau-Zener theory and Fermi's Golden Rule with an $F=1$ spin system
in a dressed atom trap with an underlying linear magnetic field
gradient.

The Landau-Zener model in general overestimates non-adiabatic
transitions, particularly for the ground state. However, for this
(good result) it is essential to use the correct speed in the
Landau-Zener expression \eqref{eq:LZ-adiabatic-limit} as discussed in
Sec.~\ref{sec:landau-zener-theory}.  For practical purposes it is
satisfactory to use the simpler Landau-Zener expressions, where
possible, and be able to err on the side of safety.

The basic results for $\Gamma/\wosc$ can be expressed in terms of a
single dimensionless parameter $\oldalpha$ for the horizontal trapping
model, or in terms of two parameters $\oldalpha$ and $\epsilon$ for
the vertical trapping model.  In the fuller and richer vertical model,
oscillatory behaviour is seen in the Fermi Golden rule treatment for
the decay rate as a function of both magnetic field gradient and Rabi
frequency. As a result, and counter to intuition with the
Landau-Zener model, higher energy states do not necessarily lead to
higher decay rates at all places in the parameter space.

In conclusion, we believe that the analytic results and procedures
presented here will be useful in the design and testing of atom traps
based on adiabatic potentials. In particular, the analytic approximate
expression given at Eq.~(\ref{Gamman}) gives a good estimate of the
expected loss rate and is easily calculated even for large values of
$\eta$, where the exact formula Eq.~(\ref{IntDecayMG}) is harder to
compute numerically.  More generally, Fig.~\ref{fig:LZOmegaBGrad}
gives an indication, in terms of lab-based parameters rather than
dimensionless variables, of the parameter region to avoid based on the
analysis and approximations used here when applied to ${}^{87}$Rb and
$F=1$.  The generalisation to higher $F$ and relaxation of some of the
approximations listed above would be useful in future work.

\begin{acknowledgments}
  The authors would like to thank Germ\'an Sinuco-Leon for helpful
  discussions and careful reading of the manuscript. This work was
  supported by the Leverhulme Trust and by EPSRC grants EP/I010394/1
  and EP/M005453/1.  We acknowledge financial support from ANR project
  SuperRing (ANR-15-CE30-0012-01). LPL is a member of Institut
  Francilien de Recherche sur les Atomes Froids (IFRAF).
\end{acknowledgments}

\appendix

\section{Effect of non-adiabatic potentials}
\label{sec:general-spin-formulae}

In this Appendix we present again the key steps of the development of
the decay rates for horizontal and vertical trapping, but this time we
keep the effect of a small non-adiabatic potential which is presented
as the dotted lines for $\Gamma_n$ in Figs.~\ref{fig:fig2},
\ref{fig:LZOmega} and~\ref{fig:LZOmegaBGrad}.  We start by considering
the Hamiltonian~\eqref{Hamiltonian}, which is already in the adiabatic
basis. As explained in Sec.~\ref{sec:quantum-dynamics-1d-trap} this
can be slightly rearranged to give Eq.~\eqref{FHamiltonian} with two
$\hat V_B$ terms: one which couples states which have a difference in
$\mfp$ of two units (proportional to $ {\hat{F}_+}^2 + {\hat{F}_-}^2$)
and one term which does not change $\mfp$ and which is proportional to
$ {\hat{F}}^2 - {\hat{F}_z}^2 $.  It is the effect of the latter term
which we focus on in this Appendix. As it will not change $\mfp$ the
term will cause a spatially dependent energy shift in the adiabatic
potentials, but only in the presence of non-adiabatic loss. This makes
the observation of this non-adiabatic potential very challenging as
its effects are only seen significantly when atoms are lost quickly
from an adiabatic trap. Nevertheless, similar kinds of effects are
discussed in the context of three-level Raman systems in
Ref.~\cite{Lacki2016}.

Thus, neglecting the $\hat V_B$ term which changes $\mfp$ in
Eq.~\eqref{FHamiltonian}, but keeping the $\hat V_B$ energy shift term
proportional to $ {\hat{F}}^2 - {\hat{F}_z}^2 $ we start with the Hamiltonian
\begin{equation}
	\hat{H} = \frac{{\hat{p}_z}^2}{2\mass} + \hat{V}_A \hat{F}_y +
        \frac{\hat{V}_B}{2} \left( {\hat{F}}^2 - {\hat{F}_z}^2 \right)
        + 
\sqrt{\Rzero^2 + {\delta(\hat{z})}^2} \hat{F}_z + \mass g \hat{z}
\label{eq:HFull-Appendix}
\end{equation}
such that the contribution to the adiabatic potentials from the $\hat
V_B$ term is considered.  As in
Sec.~\ref{sec:quantum-dynamics-1d-trap} we consider that we have a
kinetic term $\hat p_z^2/(2\mass)$, a perturbative term $\hat{V_A}
\hat{F_y}$, which is used for the coupling in Fermi's Golden rule, and
a remaining part which forms the adiabatic potentials and which now
includes a $\hat V_B$ term proportional to $ {\hat{F}}^2 -
{\hat{F}_z}^2 $.  Because of the presence of the $\hat V_B$ term in
Eq.~\eqref{eq:HFull-Appendix} the adiabatic potentials
\eqref{eq:bmg-adia-pots} are now replaced by
\begin{equation}
 V_{\mfp}(z) =
\hbar \mfp \sqrt{\Rzero^2 + \delta^2(z)} + \xxi  \frac{\Rzero^2 {\delta^\prime(z)}^2}{{\left( \Rzero^2 + \delta^2(z) \right)}^2} + \mass g z, \label{VAppendix}
\end{equation}
where in the linear regime we have a uniform gradient $\delta'$,
Eq.~\eqref{eq:delta-linearised}.  As in Sec.~\ref{RabiFrequency} the
sign term $s=g_F/\left| g_F \right|$ is absorbed into $\mfp$ and the
dressed spin states which correspond to trapping potentials are
defined to have positive $\mfp$ values. We have introduced a
parameter
\begin{equation}
 \xxi  = \frac{\hbar^2 [F(F + 1)-\mfp^2] }{4 \mass}, 
\label{eq:def:Xi}
\end{equation}
which characterises the scale of the new contribution from $\hat{V}_B
{\hat{F}_y}^2$ to the adiabatic potentials for a particular atomic
species. The parameter $\xxi$ is strictly a function of $\mfp$ (and
$F$), but as we consider here the particular case $F=1$ with initial
$\mfp=1$ and final $\mfp=0$ there will be two values of $\xxi$ which
play a role in this Appendix: $\xxi_0=\hbar^2 / ( 2 \mass)$ and
$\xxi_1=\xxi_0/2$.  The remaining part of the $\xxi$ term in
Eq.~\eqref{VAppendix} comes from $\hat V_B$, as given in
Eq.~\eqref{VBdelta}. Thus the term that multiplies $ \xxi $ in
equation \eqref{VAppendix} adds a small positive contribution to all
adiabatic potentials, regardless of dressed spin state, in the
vicinity of the resonance location. The effect of this contribution
can be reasonably ignored in the limit $\eta \rightarrow \infty$, the
limit in which cold atom traps favourably operate. However,
Figs.~\ref{fig:fig2}, \ref{fig:LZOmega} and~\ref{fig:LZOmegaBGrad}
show that there is a noticeable effect on the decay rates when we
enter deeply into the non-adiabatic region.

\subsection{Non-adiabatic potentials and the horizontal trapping model}
\label{sec:mg-model-general}

In the \mg\ model we do not need to consider the effect of
gravitational potential energy and we can express the trapping
potential as a harmonic oscillator centred around the resonant
detuning location, such that the trapping potential is given by
\begin{equation}
	V_i(z) = \hbar \Rzero +  \frac{\xxi_1}{{\oldell}^2} + \frac{1}{2} \mass \wosc^2 z^2. 
\end{equation}
The trap frequency is altered by the $\hat V_B$ contribution such that it is now given by,
\begin{equation}
	\wosc =  \sqrt{\frac{\hbar \alpha^2}{\mass \Rzero} - \frac{4
            \xxi_1}{\mass {\oldell}^4} }, 
\label{MGHumpFreq}
\end{equation}
or equivalently in terms of $\oldalpha = \oldell/\oldsigma$ and for $F=1$
\begin{equation}
 \wosc = \frac{\Rzero}{\oldalpha^2} {\left(1 + \frac{1}{\oldalpha^4}\right)}^{-1/2}. 
\label{MGHumpFreq2}
\end{equation}
Here $\oldsigma = \sqrt{\hbar/(\mass \wosc)}$ is defined using the
modified trap frequency given in Eq.~\eqref{MGHumpFreq}.  The
contribution of the $\xxi$ term could turn the curvature of the
harmonic oscillator potential negative setting a lower limit for
acceptable Rabi frequencies: $\Rzero > [ 4 {\alpha}^2
\xxi_1/\hbar]^{\frac{1}{3}} $. This gives $\Rzero / 2\pi > 1.8$~kHz
for $B^\prime = 1$~T/m and $\Rzero / 2\pi > 5.2$~kHz for $B^\prime =
5$~T/m for the state $|1,1\rangle$ of ${}^{87}$Rb.

Following the same approach as in Sec.~\ref{sec:high-gradient-model}
the wavefunction $\Phi_n(z)$ for a trapped atom is given by
Eq.~(\ref{eq:harmWF}) with the modified $a_z$, and $n$ selects the
energy of the trapped atom from the allowed energy levels which are
given by the modified expression $ E_{n} = \left( n + \frac{1}{2}
\right) \hbar \wosc + \hbar 
\Rzero + \xxi_1/\oldell^2$.

For the untrapped $ | F, \mfp = 0 \rangle $ state, i.e.\ in the absence of
gravity, the potential for the atoms is an infinite square well is
given by 
\[ V_f(z) = \left\{
  \begin{array}{l l}
     \infty, & \quad z \leq -\frac{L}{2}, \\
      \frac{\xxi_0}{{\oldell}^2}, & \quad -\frac{L}{2} \ll z \ll \frac{L}{2}, \\
	\infty, & \quad z \geq \frac{L}{2}. 
  \end{array} \right.
\]

The wavefunction of the final state is given by $ \psif =
\frac{1}{\sqrt{2 L}} \left[ e^{i k z} - {(-1)}^{\nf} e^{-i k z}
\right] \label{FullBoxWF}$ (as in Eq.~\eqref{BoxWF}) with discrete
energy levels $ E_{k(n)} = \frac{\hbar^2 k^2(n)}{2 \mass} +
\frac{\xxi_0}{{\oldell}^2}$.  The wave number $k(n)$ is altered by the
inclusion of the $\hat V_B$ contribution such that
\begin{eqnarray}
q(n) &=& k(n) \oldsigma = \sqrt{1 + 2n + 2 \Rzero/\wosc - 1/(2
  \oldalpha^2)} \nonumber\\
 &=& \sqrt{1 +2n + 2 \oldalpha^2 +     5      /(2 \oldalpha^2)}
\end{eqnarray}
(which can be compared to Eq.~\eqref{eq:q.def}).  Then the Fermi's
Golden Rule decay rate for an atom with energy $E_n$ in the \mg\ model
is given by Eq.~\eqref{IntDecayMG} with the replacements described
above for $\wosc, \oldsigma, \eta$ and $q(n)$, which are all affected
by the inclusion of the $\xxi$ term in the model.

It then follows that the analytical solution for the ground state
decay rate $\Gamma_0$ is given by Eq.~\eqref{transboxho} but with
modified variables $q_0=q(0)$, $\oldalpha$ and $\oldsigma$ and the
approximate decay rate for higher energy trapped atoms with $n > 0$ is
the similarly modified Eq.~\eqref{Gamman}.

\subsection{Non-adiabatic potentials and the vertical trapping model}
\label{sec:fg-model-general}

In the vertical trapping model case we must keep the gravitational
term in the adiabatic potential~\eqref{VAppendix} and again make a
harmonic approximation about the minimum point. When we include the
$\xxi$ term there is the complication of the additional dependence on
$\delta(z)$, even when $\delta(z)$ is linearised.  Thus, we again take
$V_i(z) = V_0 + \frac{1}{2} \mass {\wosc}^2 {\left( z -z_0 \right)}^2
$ where $z_0$ is at the centre of the displaced atom cloud which can
be determined from $\frac{d}{d z} V_{\mfp{=}1}(z) \big|_{z_0} = 0$.
As a result of the new terms in the adiabatic potential the value of
$z_0$ differs from Eq.~\eqref{eq:z0}. The harmonic potential now has
energy offset
\begin{equation}
	V_0 = \hbar \sqrt{\Rzero^2 + \delta_0^2} + \xxi_1 \frac{\Rzero^4}{{\oldell}^2 {\left( \Rzero^2 + \delta_0^2 \right)}^2} - \hbar \epsilon \delta_0 
\,,
\end{equation}
where $\delta_0 $ is the detuning at the centre of the displaced atom
cloud at $z_0$. It also has a trap frequency
\begin{equation}
	\wosc = \sqrt{\frac{\Rzero^4}{\mass {\oldell}^2} \left[ \frac{\hbar}{{\left( \Rzero^2 + \delta_0^2 \right)}^{\frac{3}{2}}} + \frac{4 \xxi_1 \Rzero^2}{{\oldell}^2} \frac{\left( 5 \delta_0^2 - \Rzero^2 \right)}{{\left( \Rzero^2 + \delta_0^2 \right)}^4} \right]}\,,
\label{eq:AppendixFinal-wosc}
\end{equation}
which cannot be written in terms of $\eta$ and $\epsilon$ alone
because the value of $\delta_0$ must be found numerically as described
above. All these changes to the potential of the initial state modify
the initial wave-function $\psii$ used in the overlap integral
Eq.~\eqref{integral-overlap-gravity}.

The potential for an untrapped atom is now approximated by a modified
Eq.~\eqref{Vslope}:
\[ V_f(z) = \left\{
  \begin{array}{l l}
     \infty, & \quad z \leq -\oldGravityBoxL, \\
    \frac{\xxi_0}{{\oldell}^2} + \mass g z, & \quad z > -\oldGravityBoxL. \\
  \end{array} \right.\]   
The stationary Schr\"odinger equation for the untrapped state can be
written as 
\begin{equation}
	E'_\kappa \, \psifa = - \frac{\hbar^2}{2 \mass} \frac{d^2 \psifa}{d z^2} + \mass g z \psifa
\end{equation}
where the $E'_\kappa$ is the modified $E_\kappa$ of Eq.~\eqref{SEfg}
and is given by $ E'_\kappa = E_\kappa - \xxi_0/\oldell^2$.

The untrapped state wavefunction is once again given by $\psifa =
\AiryConstant \mathrm{Ai} \left( \zeta \right) = \AiryConstant
\mathrm{Ai} \left[\left( z - z_\kappa \right)/ \ell \right]$, but
where now $ z_\kappa = E'_\kappa/(\mass g)$. The density of states and
normalisation co-efficient are calculated as before such that
${|\AiryConstant|}^2 \cdot D \left( E'_\kappa\right ) = 1/(\mass g
\ell^2) $.  Thus with the modified $\psifa$ and the modified
wavefunction $\psii$ described above we can obtain a modified matrix
element from Eq.~\eqref{integral-overlap-gravity}: $ \left(\sqrt{2}
  \hbar / (2 i)\right) \int_{ - \infty}^{\infty} \psiicc \hat{V}_A
\psifa \,dz $.  Then when we apply Fermi's Golden rule we find that
the decay rate $\Gamma_n$ is given by a modified
Eq.~\eqref{IntDecayFG}, but with the further modified parameters
$\beta = \oldsigma/\ell$, $\wosc$, $\oldsigma$ and $z_0$.  These
parameters were all affected by the inclusion of the $\hat V_B$
contribution to the adiabatic potentials in Eq.~\eqref{VAppendix}.

\bibliographystyle{apsrev4-1}
\bibliography{biblioBD}

\end{document}